\documentclass{sn-jnl}%

\usepackage{graphicx}%
\usepackage{multirow}%
\usepackage{amsmath,amssymb,amsfonts}%
\usepackage{amsthm}%
\usepackage{import}
\usepackage[title]{appendix}%
\usepackage{xcolor}%
\usepackage{textcomp}%
\usepackage{manyfoot}%
\usepackage{booktabs}%
\usepackage{algorithm}%
\usepackage{algorithmicx}%
\usepackage{algpseudocode}%
\usepackage{listings}%
\usepackage{bm}
\usepackage{mathrsfs}%
\usepackage{subfigure}
\usepackage{overcite}
\usepackage{footnpag}
\usepackage{soul}%
\usepackage{tikz}
\usepackage{pgfplots}
\usepackage{xcolor}
\usepackage[nohyperlinks]{acronym}
\usepackage{svg}
\usepackage{standalone}%
\usepackage{adjustbox}

\pgfplotsset{compat=newest}
\usetikzlibrary{calc}
\usetikzlibrary{positioning}%
\usetikzlibrary{shapes}%
\tikzset{>=latex} %

\newif\ifimporttikz
\newif\ifarxiv
\arxivtrue

\ifarxiv
  \importtikzfalse
\else
  \usetikzlibrary{external}
  \importtikztrue
\fi

\makeatletter
\newcommand{\importsvg}[2][]{
  \ifarxiv
    \setkeys{Gin}{#1}%
    \ifx\Gin@ewidth\relax\else\def\svgwidth{\Gin@ewidth}\fi
    \import{svg-inkscape/}{#2_svg-tex.pdf_tex}
  \else
    \includesvg[#1]{Figures/#2}
  \fi
}
\makeatother

\newif\ifshownotes
\shownotestrue %

\DeclareMathOperator*{\argmax}{arg\,max}

\acrodef{alodt}[ALoDT]{adaptive level of detail transform}
\acrodef{cvm}[CvM]{Cram\'{e}r von Mises}
\acrodef{ekf}[EKF]{extended Kalman filter}
\acrodef{elk}[ELK]{expected likelihood kernel}
\acrodef{gm}[GM]{Gaussian mixture}
\acrodef{ise}[ISE]{integral squared error}
\acrodef{fos}[FOS]{first-order stretching}
\acrodef{lam}[LAM]{likelihood agreement measure}
\acrodef{lincov}[LinCov]{linear covariance}
\acrodef{kkt}[KKT]{Karush–Kuhn–Tucker}
\acrodef{kl}[KL]{Kullback-Leibler}
\acrodef{mcr}[MCR]{maximal covariance ratio}
\acrodef{MaDEM}[MaDEM]{Mahalanobis distance of the error in the mean}
\acrodef{nise}[NISE]{normalized integral squared error}
\acrodef{nrho}[NRHO]{near rectilinear halo orbit}
\acrodef{pdf}[pdf]{probability density function}
\acrodef{sadl}[SADL]{statistical and deterministic linearization}
\acrodef{safos}[SA-FOS]{spherical-average first-order stretching}
\acrodef{sasos}[SA-SOS]{spherical-average second-order stretching}
\acrodef{sos}[SOS]{second-order stretching}
\acrodef{solc}[SOLC]{second-order linearization change}
\acrodef{stm}[STM]{state transition matrix}
\acrodef{stt}[STT]{state transition tensor}
\acrodef{sut}[SUT]{scaled unscented transform}
\acrodef{ukf}[UKF]{unscented Kalman filter}
\acrodef{us}[US]{uncertainty-scaled}
\acrodef{usfos}[US-FOS]{uncertainty-scaled first-order stretching}
\acrodef{ussolc}[US-SOLC]{uncertainty-scaled second-order linearization change}
\acrodef{wsasos}[W-SA-SOS]{whitened spherical-average second-order stretching}
\acrodef{wussadl}[W-US-SADL]{whitened uncertainty-scaled statistical and deterministic linearization}
\acrodef{wusfos}[W-US-FOS]{whitened uncertainty-scaled first-order stretching}
\acrodef{wussolc}[W-US-SOLC]{whitened uncertainty-scaled second-order linearization change}
\acrodef{wussos}[W-US-SOS]{whitened uncertainty-scaled second-order stretching}

\theoremstyle{thmstyleone}%

\theoremstyle{thmstyletwo}%

\theoremstyle{thmstylethree}%

\newlength\figureheight
\newlength\figurewidth

\begin{document}

\title[Article Title]{Unscented and Higher-Order Linear Covariance Fidelity Checks and Measures of Non-Gaussianity}

\author*[1]{\fnm{Jackson} \sur{Kulik}}\email{jackson.kulik@usu.edu}

\author[1]{\fnm{Braden} \sur{Hastings}}

\author[2]{\fnm{Keith A.} \sur{LeGrand}}

\affil[1]{\orgdiv{Mechanical and Aerospace Engineering Department}, \orgname{Utah State University}, \orgaddress{\street{1600 Old Main Hill}, \city{Logan}, \postcode{84322}, \state{Utah}, \country{USA}}}

\affil[2]{\orgdiv{School of Aeronautics and Astronautics}, \orgname{Purdue University}, \orgaddress{\street{701 W Stadium Ave}, \city{West Lafayette}, \postcode{47907}, \state{Indiana}, \country{USA}}}

\abstract{Linear covariance (LinCov) techniques have gained widespread traction in the modeling of uncertainty, including in the preliminary study of spacecraft navigation performance.
While LinCov methods offer improved computational efficiency compared to Monte Carlo based uncertainty analysis, they inherently rely on linearization approximations.
Understanding the fidelity of these approximations and identifying when they are deficient is critically important for spacecraft navigation and mission planning, especially when dealing with highly nonlinear systems and large state uncertainties.
This work presents a number of computational techniques for assessing linear covariance performance.
These new LinCov fidelity measures are formulated using higher-order statistics, constrained optimization, and the unscented transform.}


\maketitle

\section{Introduction}
Uncertainty propagation in astrodynamics underlies satellite navigation, space situational awareness, and robust/uncertainty-aware trajectory optimization.
Often in these applications, theoretical and practical constraints motivate the usage of linearization and/or Gaussian assumptions in the approximation of propagated uncertainty.
These constraints include limited computational resources in autonomous systems and, in other cases, the need for semi-analytical or convex formulations of uncertainty propagation.
\Ac{lincov} analysis is based around the same principles as the prediction step in the extended Kalman filter and employs first-order Taylor approximations of the dynamics to obtain a closed-form covariance propagation \cite{carpenter2025navigation, geller2006linear}.
The validity of this approach relies on the approximate linearity of the dynamics within the effective support of the distribution, which is commonly assumed to be Gaussian.
Other uncertainty propagation methods that handle weak nonlinearity while assuming Gaussianity of the distribution include the unscented transform \cite{julier2004unscented}.
Still other nonlinear non-Gaussian uncertainty propagation methods include Monte Carlo, conjugate unscented transform \cite{adurthi2012conjugate}, state transition tensor-based moment propagation \cite{park2006nonlinear}, Gaussian mixture propagation \cite{alspach1972NonlinearBayesianEstimation,demars2013entropy}, differential algebra \cite{servadio2022maximum}, and polynomial chaos expansion approaches \cite{jones2013nonlinear}.
Given \ac{lincov}'s simplicity, speed, and overall adoption by practitioners, it is likely to still play a large role in future spacecraft mission planning.
Thus, it is important to establish fast fidelity checks to accompany this technique to indicate whether the outputs are trustworthy or that more sophisticated nonlinear uncertainty propagation methods are needed.

Thus, two related questions arise in measuring the fidelity of a \ac{lincov} approximation: how non-Gaussian is a propagated distribution, and how nonlinear are the dynamics on the effective support of the distribution?
To address the question of quantifying non-Gaussianity, Mercurio et al.\ demonstrated how moment-based approaches, information theoretic approaches including \ac{kl}-divergence, and shape-based metrics such as the Wasserstein distance can be employed to assess the non-Gaussianity of a distribution \cite{mercurio2018HowNongaussian}.
On the other hand, Junkins presented a methodology for assessing the degree of nonlinearity of the flow of a dynamical system. \cite{junkins2004nonlinear}
Recent work has improved on methods to calculate that nonlinearity index using \acp{stt}\cite{jenson2023semianalytical, kulik2024applications}.
Other recent work studied measures of linearized uncertainty realism through the lens of the nonlinearity index \cite{gutierrez2024classifying}.
Work on splitting in adaptive Gaussian mixture uncertainty propagation has considered uncertainty weighted measures of nonlinearity \cite{kulik2025NonlinearityUncertaintySplitting,zanetti2025uncertainty,calkins2024efficient,calkins2025dynamics,siciliano2025higher,iannamorelli2025AdaptiveGaussianMixture}.
Additionally, recent work on covariance steering for robust/uncertain trajectory optimization has enforced the fidelity of the approximations employed in computations by leveraging nonlinearity indices in a statistical context \cite{fife-steering,qi2025optimal}.
Other approaches have quantified non-Gaussianity in terms of skewness in Gaussian mixture splitting \cite{dunik2018directional,horwood2011adaptive} and covariance steering settings\cite{qioptimal} often employing a scalar summary metric of overall skewness using tensor methods from higher-order statistics \cite{mccullagh2018tensor}.
This work summarizes, extends, and formalizes some of these approaches to uncertainty, nonlinearity, and non-Gaussianity quantification.

In particular, this work presents a number of novel moment-based and constrained optimization-based approaches for quantifying both non-Gaussianity and the fidelity of the linearization assumption in \ac{lincov} analysis.
Where possible, we present multiple methods for calculating each metric that are based either on state transition tensor or unscented transform methodologies.
Using higher-order approximations of the mean and covariance, we present multiple methods for assessing the approximation quality of linear approximation-based moment propagation.
These measures are based on expected Mahalanobis distances and maximal differences in the marginal variances between two fidelities of approximation.
In the area of higher-order statistics we present a new connection between standard multivariate Gaussian kurtosis and the fourth-order identity tensor.
Based on that observation, we offer a new compact definition of the excess kurtosis tensor that offers a novel method of standardization for the multivariate case.
We then present and demonstrate how to compute novel scalar summary statistics for the maximal standardized directional skewness and kurtosis.

The remainder of this article is arranged as follows.
Background information on partial derivative tensors, whitening transformations, and statistical moments is given in Section~\ref{sec:background}.
Section~\ref{sec:fidelity_checks} presents novel expectation-based measures of fidelity and their Monte Carlo, unscented, and second-order approximations.
Fidelity measures based on constrained optimization problems are discussed in~\ref{sec:optimization_based_fidelity_measures}.
Section~\ref{sec:applications} demonstrates the application of these methods to a cislunar orbit \ac{lincov} analysis, and conclusions are drawn in Section~\ref{sec:conclusion}.

\section{Background}
\label{sec:background}
Consider a nonlinear function $\mathbf{g}:\mathbb{R}^n\rightarrow\mathbb{R}^m$ that may represent the flow of a dynamical system or a measurement function.
Let $\mathbf{x}\in\mathbb{R}^n$ be a random variable with mean $\boldsymbol{\mu}_x$ and covariance $\mathbf{P}_x$.
Linear covariance techniques rely on the following simple principle for uncertainty propagation:
for sufficiently small input covariance $\mathbf{P}_x$, the resulting output mean and covariance are approximately
\begin{align}
\textrm{E}[\mathbf{z}] \approx
    \boldsymbol{\mu}_z^{(1)}&=\mathbf{g}(\boldsymbol{\mu}_x)\label{eq:lin_mean}\\
    \mathbf{P}_{z} \approx
    \mathbf{P}_z^{(1)}&=\mathbf{G}\mathbf{P}_x\mathbf{G}^T
    \label{eq:lin_cov}
\end{align}
where the superscript ``$(1)$'' denotes that these quantities arise from linear covariance analysis and
\begin{equation}
    \mathbf{G}=\frac{\mathrm{d}\mathbf{g}}{\mathrm{d}\mathbf{x}}\Big\vert_{\boldsymbol{\mu}_x}
\end{equation}
is the Jacobian of $\mathbf{g}$ evaluated at $\boldsymbol{\mu}_x$.
Furthermore, if $\mathbf{x}$ is Gaussian-distributed, then by the same aforementioned assumptions, $\mathbf{z}$ is also approximately Gaussian distributed.
Often, linear covariance techniques are employed for stochastic systems that include  process noise or measurement noise.
In this work, we will examine only deterministic systems, where the dominating source of output uncertainty is due the initial uncertainty in $\mathbf{x}$.

This paper develops new methods and metrics for analyzing the quality of LinCov-estimated quantities in nonlinear systems.
The methods in this work are based primarily on calculating expectations related to linearization error or solving constrained optimization problems related to linearization error.
In both cases, the proposed methods leverage higher-order statistics or the unscented transform.
For multivariate random variables, central moments beyond order two require tensor representations.
Thus, our constrained optimization approach leverages the theory of tensor eigenvalues and tensor operator norms \cite{jenson2023semianalytical,kulik2024applications}.
One of the most important elements of our approach is output-whitening, which enables each output to be meaningfully interpretable as a Mahalanobis distance and is unit and scale-invariant.

\subsection{Partial Derivative Tensors}
\label{sec:partial_deriv_tensors}
In this work, we will heavily rely on higher-order partial derivative tensors of the nonlinear function $\mathbf{g}$ in order to describe the higher-order terms in the Taylor series expansion of $\mathbf{g}$ about the mean of $\boldsymbol{\mu}_x$ as these will approximately describe the error in approximating $\mathbf{g}(\boldsymbol{\mu}_x+\delta\mathbf{x})$ by its linearization $\mathbf{g}(\boldsymbol{\mu}_x)+\mathbf{G}\delta\mathbf{x}$.
The Taylor series expansion of $\mathbf{g}$ about the mean $\boldsymbol{\mu}_x$ is given by
\begin{equation}
    \mathbf{g}(\boldsymbol{\mu}_x+\delta\mathbf{x})=\mathbf{g}(\boldsymbol{\mu}_x)+\mathbf{G}\delta\mathbf{x}+\frac{1}{2}\frac{\partial^2 \mathbf{g}}{\partial \mathbf{x}^2}\Big\vert_{\boldsymbol{\mu}_x}\delta\mathbf{x}^2+\mathcal{O}(\delta\mathbf{x}^3)
\end{equation}
where we define the shorthand notation above for double contraction of the tensor with two copies of a vector in terms of the following Einstein notation (where there is a sum over any repeated indices in an expression)
\begin{equation}
    \left(\frac{\partial^2 \mathbf{g}}{\partial \mathbf{x}^2}\Big\vert_{\boldsymbol{\mu}_x}\delta\mathbf{x}^2\right)^{i}=\left(\frac{\partial^2 \mathbf{g}}{\partial \mathbf{x}^2}\Big\vert_{\boldsymbol{\mu}_x}\right)^i_{j,k}\delta\mathbf{x}^j\delta\mathbf{x}^k
\end{equation}
In the following, we adopt the abbreviated notation
\begin{equation}
    \mathbf{G}^{(2)}=\frac{\partial^2 \mathbf{g}}{\partial \mathbf{x}^2}\Big\vert_{\boldsymbol{\mu}_x}
\end{equation}
to denote the second-order partial derivative tensor. We will primarily be concerned with partial derivatives associated with the flow of a dynamical system. The partial derivatives of the flow of a dynamical system are known as state transition tensors and can be computed by deriving and integrating variational equations associated with the dynamical system \cite{park2006nonlinear} or by using differential algebra techniques \cite{rasotto2016differential}.

\subsection{Whitening and Mahalanobis Distance}
Given a random variable $\mathbf{x}$ and its covariance matrix $\operatorname{var}(\mathbf{x})=\mathbf{P}$, a whitening transform is defined by any transformation
\begin{align}
    \mathbf{y} = \mathbf{W} \mathbf{x}
\end{align}
such that $\mathbf{y}$ is the same dimension of $\mathbf{x}$ and the ``whitened'' covariance matrix $\operatorname{var}(\mathbf{y}) = \bm{I}$.
It is straightforward to show that the inverse of any square root factor $\mathbf{W} = \mathbf{P}^{-1/2}$, where $\mathbf{P} = \mathbf{P}^{1/2} (\mathbf{P}^{1/2})^{T}$ provides a whitening transformation.
A convenient property of the whitened vector $\mathbf{y}$ is that its 2-norm is equal to the Mahalanobis distance in the original space:
\begin{align}
    \sqrt{\mathbf{y}^T \mathbf{y}} = \sqrt{\mathbf{x}^{T} \mathbf{P}^{-1} \mathbf{x}}
\end{align}

\subsection{Moments}
\label{sec:moments}
Consider a scalar random variable $x$ with density function $p(x)$.
The $m$\textsuperscript{th} central moment of $x$, if it exists, is given by
\begin{align}
    \mathrm{E}[(x - \mathrm{E}[x])^{m}] = \int (x - \mathrm{E}[x])^{m} p(x) \mathrm{d}
x\end{align}
For a multivariate random variable $\mathbf{x}$ with density $p(\mathbf{x})$, one notion of directional moments is in terms of a scalar central moment of the one-dimensional marginal distribution along a given direction $\hat{\mathbf{v}}$.
This marginalization is performed over $V_{\perp}(\hat{\mathbf{v}})$, the subspace of the domain of $\mathbf{x}$ that is orthogonal to $\hat{\mathbf{v}}$.
That is, the directional marginal is
\begin{align}
    p(\alpha)= \int_{V_{\perp}(\hat{\mathbf{v}})} p(\mathbf{x}) \mathrm{d}V^{n-1}
\end{align}
where $\mathbf{x}=\alpha \hat{\mathbf{v}} + \mathbf{u}$ for some $\mathbf{u} \in V_{\perp}(\hat{\mathbf{v}})$ and $\mathrm{d}V^{n-1}$ is an $n-1$ dimensional differential volume.
Then, the $m$\textsuperscript{th} \textit{marginal central moments} are defined as
\begin{align}
    \mathrm{E}[(\alpha - \mathrm{E}[\alpha])^m]
\end{align}
where the expectations are understood to be taken with respect to the marginal distribution $p(\alpha)$.
The integrals involved in the marginalization and the expectation over $\alpha$ can be combined to express the $m$\textsuperscript{th} $\hat{\mathbf{v}}$-marginal central moment compactly as
\begin{align}
     \mathrm{E}[(\alpha - \mathrm{E}[\alpha])^m]
     = \mathrm{E}\left[\left(\hat{\mathbf{v}}^{T}(\mathbf{x} - \bm{\mu}_{x})\right)\right]
\end{align}
The \textit{standardized} marginal central moments of the distribution are given as
\begin{equation}
    \label{eq:standardized_marginal_central_moments}
    \frac{\mathrm{E}[(\alpha - \mathrm{E}[\alpha])^m]}{\mathrm{E}[(\alpha - \mathrm{E}[\alpha])^2]^{m/2}}=\frac{\mathrm{E}[\left(\hat{\mathbf{v}}^{T}(\mathbf{x} - \bm{\mu}_{x})\right)^m]}{\mathrm{E}[\left(\hat{\mathbf{v}}^{T}(\mathbf{x} - \bm{\mu}_{x})\right)^2]^{m/2}}
\end{equation}

Having discussed marginal higher-order central moments, we will introduce higher-order central moment tensors and highlight related and distinct notions about directional information that can be obtained from these. Denote the linearly propagated distribution of $\mathbf{x}$ by
\begin{equation}
    \mathbf{z}^{(1)}=\mathbf{g}(\boldsymbol{\mu}_x)+\mathbf{G}(\mathbf{x}-\boldsymbol{\mu}_x)
\end{equation}
If $\mathbf{x}$ is Gaussian distributed, then so too is $\mathbf{z}^{(1)}$, and thus its statistics are described completely by the mean and covariance in~\eqref{eq:lin_mean} and~\eqref{eq:lin_cov}, respectively.
The third-order central moments of this Gaussian distribution along any direction are zero, so
\begin{equation}
    (\mathbf{S}_z^{(1)})^{i,j,k}=\mathrm{E}[(\mathbf{z}^{(1)}-\boldsymbol{\mu}_z^{(1)})^i(\mathbf{z}^{(1)}-\boldsymbol{\mu}_z^{(1)})^j(\mathbf{z}^{(1)}-\boldsymbol{\mu}_z^{(1)})^k]=0
\end{equation}
We define the standardized third-order central moment---to be referred to as the \textit{skewness} tensor---of a distribution as the third-order central moment of the random variable that results from whitening the original random variable
\begin{equation}
    \boldsymbol{\Sigma}^{i,j,k}=(\mathbf{P}^{-1/2})^{i}_{i'}(\mathbf{P}^{-1/2})^{j}_{j'}(\mathbf{P}^{-1/2})^{k}_{k'}S^{i',j',k'}
    \label{eq:skewness}
\end{equation}
which is also zero in the case of $\mathbf{z}^{(1)}$.
The fourth order central moment tensor of $\mathbf{z}^{(1)}$ is defined as
\begin{align}
        (\mathbf{K}_z^{(1)})^{i,j,k,l}&=\mathrm{E}[(\mathbf{z}^{(1)}-\boldsymbol{\mu}_z^{(1)})^i(\mathbf{z}^{(1)}-\boldsymbol{\mu}_z^{(1)})^j(\mathbf{z}^{(1)}-\boldsymbol{\mu}_z^{(1)})^k)(\mathbf{z}^{(1)}-\boldsymbol{\mu}_z^{(1)})^l]\\
        &=(\mathbf{P}_z^{(1)})^{i,j}(\mathbf{P}_z^{(1)})^{k,l}+(\mathbf{P}_z^{(1)})^{i,k}(\mathbf{P}_z^{(1)})^{j,l}+(\mathbf{P}_z^{(1)})^{i,l}(\mathbf{P}_z^{(1)})^{j,k}\\
        &=\operatorname{sym}\left(3(\mathbf{P}_z^{(1)})^{i,j}(\mathbf{P}_z^{(1)})^{k,l}\right)
\end{align}
where the second equality known as Isserlis' theorem can be obtained via the joint characteristic function\cite{park2006nonlinear, mccullagh2018tensor} or by way of a symmetry argument, inverse whitening transformation, and by counting the number of pairs that can be selected from four indices.
The $\operatorname{sym(\cdot)}$ operator denotes averaging the tensor over all permutations of the indices to symmetrize the tensor
\begin{equation}
    \operatorname{sym}(\mathbf{T})^{i_1,...,i_m}=\frac{1}{m!}\sum_{\sigma\in S_m}\mathbf{T}^{\sigma(i_1,...,i_m)}
\end{equation}
The standardized fourth order central moment tensor or \textit{kurtosis} tensor is obtained by applying a whitening transformation to $\mathbf{z}^{(1)}$ and taking the fourth central moment of the resulting random variable:
\begin{align}
    (\boldsymbol{\kappa}_z^{(1)})^{i,j,k,l}
    &=((\mathbf{P}_z^{(1)})^{-1/2})^{i}_{i'}((\mathbf{P}_z^{(1)})^{-1/2})^{j}_{j'}((\mathbf{P}_z^{(1)})^{-1/2})^{k}_{k'}((\mathbf{P}_z^{(1)})^{-1/2})^{l}_{l'}(\mathbf{K}_z^{(1)})^{i',j',k',l'}\\
    &=3(\mathbf{I}^{(4)}_m)^{i,j,k,l}
\end{align}
where $\mathbf{I}^{(4)}_m$ is the fourth-order, $m$-dimensional identity tensor \cite{kolda2011shifted,qi2005eigenvalues} %
such that
\begin{equation}
    \mathbf{I}^{(4)}_m\boldsymbol{\xi}^3=\boldsymbol{\xi} \qquad \text{and} \qquad
    \mathbf{I}^{(4)}_m\boldsymbol{\xi}^4=1
\end{equation}
for all $\boldsymbol{\xi}\in\mathbb{R}^m$ such that $\Vert \boldsymbol{\xi}\Vert_2=1$. The relationship between the Gaussian kurtosis and the identity tensor is a novel observation as far as the authors are aware. We define the \textit{excess kurtosis} tensor as the difference between the kurtosis $\boldsymbol{\kappa}$ of some distribution and the kurtosis of a Gaussian distribution of the same dimension.
That is,
\begin{equation}
    \delta\boldsymbol{\kappa}=\boldsymbol{\kappa}-3\mathbf{I}^{(4)}
    \label{eq:excess_kurtosis}
\end{equation}
The skewness and kurtosis tensors as defined here are the central moments of the whitened random variable $\mathbf{y}$ such that
\begin{equation}
    (\mathbf{y}-\boldsymbol{\mu}_y)=\mathbf{P}_x^{-1/2}(\mathbf{x}-\boldsymbol{\mu}_x)
\end{equation}
where $\mathbf{x}$ is the original random variable under consideration. Note that the skewness and excess kurtosis are both identically zero for the linearly propagated distribution as it is Gaussian, though this will not be true in general.
Note that an $m$th-order central moment tensor that is not standardized (such as $\mathbf{S}$) can be used to find the corresponding $m$th-order central moment of a marginal distribution along any particular direction (such as the skewness of the marginal distribution along $\hat{\mathbf{v}}$) by contraction $m$ times with a unit vector along that direction.
On the other hand, repeated contraction along a particular direction of the \textit{standardized} central moment tensors (skewness $\boldsymbol{\Sigma}$ and kurtosis $\boldsymbol{\kappa}$) does \textit{not} necessarily result in the standardized skewness and kurtosis of the marginal distribution under contraction (as defined in~\eqref{eq:standardized_marginal_central_moments}).
For example,
\begin{equation}
\label{eq:directional_standardized_skewness}
    \Sigma^{i,j,k}\hat{\mathbf{v}}^i\hat{\mathbf{v}}^j\hat{\mathbf{v}}^k\neq \frac{S^{i,j,k}\hat{\mathbf{v}}^i\hat{\mathbf{v}}^j\hat{\mathbf{v}}^k}{(\hat{\mathbf{v}}^T\mathbf{P}\hat{\mathbf{v}})^{3/2}}=\frac{\mathrm{E}[(\hat{\mathbf{v}}^T(\mathbf{x}-\boldsymbol{\mu})]^3}{\sigma_{\hat{\mathbf{v}}}^3}
\end{equation}
where $\sigma_{\hat{\mathbf{v}}}$ is the standard deviation of the marginal distribution along an arbitrary unit vector $\hat{\mathbf{v}}$.
We refer to the left hand side of~\eqref{eq:directional_standardized_skewness} as the \textit{direction standardized skewness}.

One case in which a directional standardized central moment and the corresponding marginal central moment are equivalent include when the distribution is a homoscedastic multivariate Gaussian.
Another special case that leads to equivalence is when the distribution is Gaussian and the direction of marginalization aligns with a principal axis of the covariance ellipsoid.
Thus, one must be careful not to conflate the standardized skewness tensor contracted along a particular direction with the marginal skewness along that direction.
Instead, the standardized skewness tensor contracted along a particular direction should be interpreted as the skewness of the marginal distribution in the given direction of the whitened random variable.

Our definition of the standardized central higher-order moment tensors yields a homogeneous polynomial representation of the higher-order moments in a manner that is independent of the location and spread of the distribution as described by mean and covariance, respectively.
By this invariance to mean and covariance, these tensors provide a standardized, interpretable, and computationally tractable description of non-Gaussianity and its overall structure in a multivariate setting.
In contrast, we expect that no marginalization-based definition of directional higher-order moments bears this important homogeneity property.
This is due to the quotient nature of the scalar standardized central moments, which generally does not afford a homogeneous polynomial description.

\subsection{Tensor Eigenvalues}
Given an $m$th-order supersymmetric (symmetric under any permutation of the indices) covariant tensor (a multilinear functional that operates on $m$ vectors to produce a scalar value) $\mathbf{T}$, consider the following constrained optimization problem:
\begin{align}
    \max_{\Vert\mathbf{x}\Vert_2=1}\mathbf{T}_{i_1,...,i_m}\mathbf{x}^{i_1}...\mathbf{x}^{i_m}
\end{align}
The method of Lagrange multipliers may be applied to this constrained optimization to yield the following conditions for optimality:
\begin{align}
    \mathbf{T}\mathbf{x}^{m-1}&=\lambda\mathbf{x}\\
    \Vert\mathbf{x}\Vert_2&=1
\end{align}
The above equations define a z-eigenvalue and eigenvector pair. The theory\cite{de2000best,lim2005singular} behind tensor z-eigenvalues and their computation\cite{kolda2006matlab,kolda2011shifted}  has recently been adopted by the astrodynamics community to study nonlinearity and to quantify the error associated with linear methods for guidance, navigation, and control \cite{jenson2022semianalytical,jenson2023semianalytical,kulik2024applications}. In order to calculate the maximal z-eigenvector pair, an algorithm known as shifted higher-order power iteration may be employed. Shifted higher-order power iteration is a generalization of shifted power iteration for finding eigenvalues of symmetric matrices. The algorithm is described by the iteration
\begin{equation}
    \mathbf{x}_{(k+1)}=\frac{\mathbf{T}\mathbf{x}_{(k)}^{m-1}+\alpha\mathbf{x}_{(k)}}{\Vert\mathbf{T}\mathbf{x}_{(k)}^{m-1}+\alpha\mathbf{x}_{(k)}\Vert_2}
\end{equation}
where the subscript $(k)$ indicates the $k$th
 iterate of the algorithm and $\alpha$ is chosen to ensure global convergence of the algorithm to an eigenpair of the tensor $\mathbf{T}$. We use the notation $\mathbf{T}\mathbf{x}^{m-1}$ as shorthand for the the following operation that produces a vector
 \begin{equation}
(\mathbf{T}\mathbf{x}^{m-1})_{i_1}=T_{i_1,i_2,...,i_m}x^{i_2}...x^{i_m}
 \end{equation}
The shift parameter $\alpha$ must be chosen to be larger than the order of the tensor minus one multiplied with the spectral radius of the matrix resulting from contracting all but two of the indices of the tensor with any vector on the unit sphere:
 \begin{equation}
\alpha>(m-1)\max_{\Vert\mathbf{x}\Vert_2=1}\rho(\mathbf{T}\mathbf{x}^{m-2})
 \end{equation}
One simple method to compute a sufficient shifting factor $\alpha$ is given by the sum of the absolute values of the entries of the tensor\cite{kolda2011shifted}
\begin{equation}
\alpha=(m-1)\sum_{i_1,...,i_m}\vert T_{i_1,...,i_m}\vert
\end{equation}
This algorithm will converge regardless of the initial guess for the eigenvector, and will tend to converge to eigenpairs with the largest eigenvalues, though is not guaranteed to converge to the eigenpair with the largest eigenvalue from every initial guess. As such, if there is not a good method for generating an initial guess at the eigenvector corresponding to the largest eigenvalue, then the algorithm should be run with a number of random initial guesses.

\section{Expectation-Based Fidelity Checks}%
\label{sec:fidelity_checks}
This section presents a variety of measures for assessing the fidelity of a given \ac{lincov} approximation.
We first propose a set of baseline measures in terms of the differences and ratios of the true and \ac{lincov}-approximated moments.
The role of higher-order statistical moments as measures of fidelity is also discussed.
Monte Carlo based approximations of these fidelity measures are then discussed and serve as a benchmark for comparative analysis.
Second-order Taylor and unscented approximations are then proposed.

\subsection{Baseline Expectation-Based Measures}
\label{sec:baseline_monte_carlo_approaches}
This subsection proposes a number of baseline measures of \ac{lincov} fidelity that are expressed in terms of the true but unknown mapped distribution $p_{z}(\mathbf{z})$ and its moments.
Following subsections then detail different methods by which these measures may be obtained using Monte Carlo, higher-order polynomial, and unscented approximations.

\begin{figure}[htbp]
    \centering
  
  \ifimporttikz
    \tikzsetnextfilename{md_means_illustration}
    \import{Figures/}{md_means_illustration.tikz}
  \else
    \includegraphics{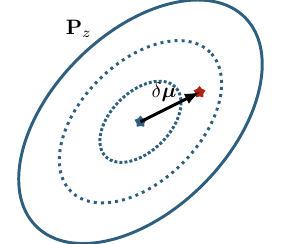}
  \fi

\caption{Notional depiction of the contours of equal Mahalanobis distance with a difference in two means superimposed.}
\label{fig:mean_dif}
\end{figure}

\textbf{Squared Mahalanobis Distance of the Means}
As a simple fidelity test, we can examine the moments of the true random variable $\mathbf{z}$ and the linearly propagated random variable
\begin{equation}
    \mathbf{z}^{(1)}=\mathbf{g}(\boldsymbol{\mu}_x)+\mathbf{G}(\mathbf{x}-\boldsymbol{\mu}_x)
\end{equation}
Denote by $\bm{\mu}_{z}$ and $\mathbf{P}_{z}$ the (unkown) true mean and covariance, respectively, of the random variable $\mathbf{z}$.
The absolute difference between the true mean and the \ac{lincov} mean can be taken as a metric of interest.
Alternatively, the absolute difference can be normalized to a squared Mahalanobis distance according to
\begin{equation}
    \label{eq:mean-dif}
    \texttt{SMDM}_{\mathbf{g},\mathbf{x}}
    =
    (\boldsymbol{\mu}_z-\boldsymbol{\mu}_z^{(1)})^T(\mathbf{P}^{(1)}_z)^{-1}(\boldsymbol{\mu}_z-\boldsymbol{\mu}_z^{(1)})
\end{equation}
This measure is referred to as the \textit{squared Mahalanobis distance of the means} and is depicted in Fig.~\ref{fig:mean_dif}.

\textbf{Maximal Ratio of Moments}

\begin{figure}
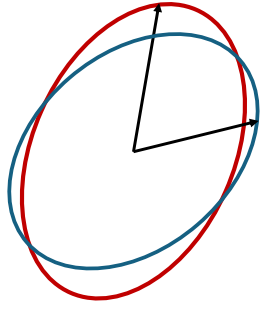

    \centering
\adjustbox{trim=0cm 0.5cm 0cm 0.5cm, clip}{%
  \importsvg[width=10cm]{ellipse_ratios1}%
}
\caption{Notional depiction of minimal and maximal ratios of two ellipses. Note that the two directions are not orthogonal.}
\label{fig:ratios}
\end{figure}

The concept of moment comparison can be extended to the second moments.
The minimum and maximum values of the generalized Rayleigh quotient
\begin{equation}
\frac{\boldsymbol{\zeta}^T(\mathbf{P}_{z}^{(1)})^{-1}\boldsymbol{\zeta}}{\boldsymbol{\zeta}^T(\mathbf{P}_{z})^{-1}\boldsymbol{\zeta}}
\end{equation}
can be employed to assess the difference between the two distributions.
This generalized Rayleigh quotient can be interpreted as the minimal/maximal linear covariance Mahalanobis distance of the 1-sigma contour sample covariance.
The minimum and maximum values of this generalized Rayleigh quotient can be calculated by solving the generalized eigenvalue problem with
\begin{equation}
\label{eq:gen_eig}
    (\mathbf{P}_{z}^{(1)})^{-1}\boldsymbol{\zeta}=\lambda (\mathbf{P}_{z})^{-1}\boldsymbol{\zeta}
\end{equation}
and finding the minimum and maximum eigenvalue which will give the minimum and maximum ratios.
The corresponding measure of fidelity is referred to as the \ac{mcr}
and given  by
\begin{align}
\label{eq:mcr_definition}
    \texttt{MCR} = \max(1/ \min_{i} \lambda_{i}, \,  \max_{i} \lambda_{i})
\end{align}
A notional depiction of the minimal and maximal ratio directions of two ellipsoids is shown in Fig. \ref{fig:ratios}.

\textbf{Expected Squared Mahalanobis Distance}

The expected squared Mahalanobis distance of a random variable $\boldsymbol{\zeta}$ with respect to the linearly propagated mean and covariance is
\begin{align}
\label{eq:esmd}
    \texttt{ESMD}_{\mathbf{g},\mathbf{x}}=\mathrm{E}[(\mathbf{z}-\boldsymbol{\mu}_z^{(1)})^T(\mathbf{P}^{(1)}_{z})^{-1}(\mathbf{z}-\boldsymbol{\mu}_z^{(1)})]
\end{align}
If the true distribution is close to the linearly propagated distribution, then the expected squared Mahalanobis distance should be given by $m$, the dimension of $\mathbf{z}$, which is the mean of a Chi-square random variable with $m$ degrees of freedom.
Eq.~\eqref{eq:esmd} can also be written in tensor notation as
\begin{equation}
    \texttt{ESMD}_{\mathbf{g},\mathbf{x}} = \mathrm{E}[ (\mathbf{P}^{(1)})^{-1}_{ij} (\mathbf{z}-\boldsymbol\mu_{z}^{(1)})^{j}  (\mathbf{z}-\boldsymbol\mu_{z}^{(1)}  )^{i} ]
\end{equation}

Because $ (\mathbf{P}^{(1)} )^{-1}_{ij}$ is deterministic, we can pull it outside the expectation to give
\begin{align}
    \texttt{ESMD}_{\mathbf{g},\mathbf{x}} &=  (\mathbf{P}^{(1)} )^{-1}_{ij} \mathrm{E} [   (\mathbf{z}-\boldsymbol\mu_{z}^{(1)} )^{j}  (\mathbf{z}-\boldsymbol\mu_{z}^{(1)} )^{i} ]\\
    &=  (\mathbf{P}^{(1)} )^{-1}_{ij} \mathrm{E} [   (\mathbf{z}-\bm{\mu}_{z} + \bm{\mu}_{z} - \boldsymbol\mu_{z}^{(1)} )^{i}  (\mathbf{z}-\bm{\mu}_{z} + \bm{\mu}_{z} - \boldsymbol\mu_{z}^{(1)} )^{j} ]\\
    \label{eq:esmd_P_factored}
    &=  (\mathbf{P}^{(1)} )^{-1}_{ij}  (\mathbf{P}_{\zeta}^{ij} + \delta\boldsymbol{\mu}^{j} \delta\boldsymbol{\mu}^{i}  )
\end{align}
where $\delta\boldsymbol{\mu} = \bm{\mu}_{z} - \boldsymbol\mu_z^{(1)}$.
The first term in \eqref{eq:esmd_P_factored} can be reformulated as
\begin{align}
 (\mathbf{P}^{(1)} )^{-1}_{ij}  (\mathbf{P}_{z} )^{ij} = \operatorname{tr} (  ( \mathbf{P}^{(1)} )^{-1}\mathbf{P}_{z})
\end{align}
which is also the sum of the generalized eigenvalues of the pair of covariance matrices. Thus, this measure can be conveniently expressed in terms of means and covariance matrices as
\begin{align}
\label{eq:ESMD}
\texttt{ESMD}_{\mathbf{g},\mathbf{x}} = \operatorname{tr} (  ( \mathbf{P}^{(1)} )^{-1}\mathbf{P}_{z}  ) + \delta\boldsymbol\mu^T (\mathbf{P}^{(1)})^{-1} \delta\boldsymbol\mu
\end{align}
and this can be computed from samples by directly employing the sample covariance and the difference between the sample mean and the propagated mean.

\begin{figure}
    \centering

  \ifimporttikz
    \tikzsetnextfilename{expected_mh_illustration}
    \import{Figures/}{expected_mh_illustration.tikz}
  \else
    \includegraphics{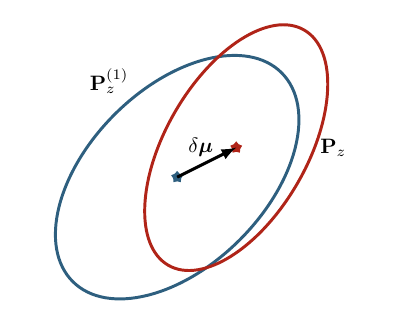}
  \fi

\caption{Notional depiction of two distributions with distinct means and covariance ellipsoids.}
\label{fig:expected_mh_graphic}
\end{figure}

The separation of the two terms in~\eqref{eq:ESMD} provides insight into the effects of the difference in means on the expected Mahalanobis distance and the effects of the difference in distribution about the respective means.
The $\operatorname{tr} (  ( \mathbf{P}^{(1)} )^{-1}\mathbf{P}_{\zeta}  )$ term in~\eqref{eq:ESMD} can be interpreted as the expected Mahalanobis distance if the means of the random variable and the linear propagation are the same.
The $ \delta\boldsymbol\mu^T (\mathbf{P}^{(1)})^{-1} \delta\boldsymbol\mu $ term is the expected Mahalanobis distance from the difference in means between the true random variable and the linear propagation.
In Fig.~\ref{fig:expected_mh_graphic}, the two different effects accounted for in expected squared Mahalanobis distance are displayed: a difference in means and a difference in covariances.
The expected square Mahalanobis distance also bears an interesting connection to information theoretic interpretations of distribution similarity.
Specifically, in the special case where the compared distributions are both Gaussian, the expected squared Mahalanobis distance is related to the \ac{kl} divergence as
\begin{align}
  \texttt{ESMD}_{\mathbf{g},\mathbf{x}} = 2 \mathrm{D}_{\textrm{KL}}(p(\mathbf{z})||p(\mathbf{z}^{(1)})) + m - \log |\mathbf{P}^{(1)} \mathbf{P}_{z}^{-1}|
\end{align}

\textbf{Expected Squared Mahalanobis Distance of Linearization Error}

Alternatively, we can examine the behavior of the linearization error itself.
Consider the first-order functional approximation of the nonlinear map given by
\begin{equation}
    \mathbf{z}^{(1)}(\mathbf{x})=\mathbf{g}(\boldsymbol{\mu}_x)+\mathbf{G}(\mathbf{x}-\boldsymbol{\mu}_x)
\end{equation}
so the expectation of the squared Mahalanobis distance of the linearization error is
\begin{align}
\label{eq:expected_squared_mahalanobis_distance_linearization_error}
  \texttt{ESMDoLE}_{\mathbf{g},\mathbf{x}}
  =
  \mathrm{E}
  \left[
    (\mathbf{g}(\mathbf{x}) - \mathbf{g}(\bm{\mu}_{x}) - \mathbf{G}(\mathbf{x} - \bm{\mu}_{x}))^{T}
    (\mathbf{P}^{(1)})^{-1}
    (\mathbf{g}(\mathbf{x}) - \mathbf{g}(\bm{\mu}_{x}) - \mathbf{G}(\mathbf{x} - \bm{\mu}_{x}))
  \right]
\end{align}

\textbf{Maximal Directional Standardized Higher-Order Moments}

\begin{figure}[htbp]
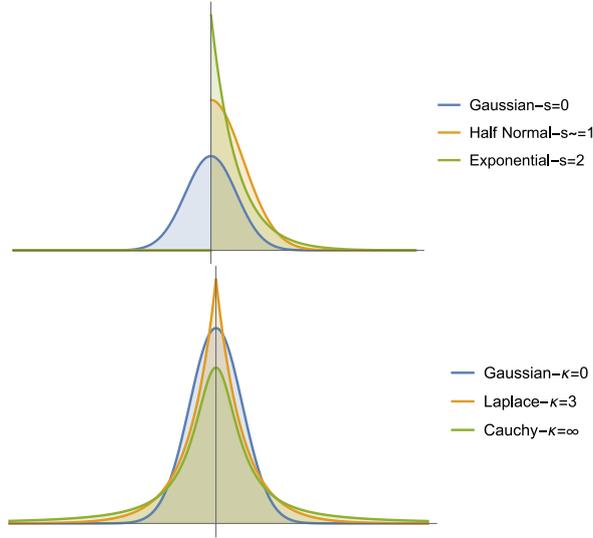

    \centering
\adjustbox{trim=1cm 0cm 0cm 0cm, clip}{%
  \importsvg[width=8.5cm]{skewness_demo}%
}
\adjustbox{trim=1cm 0cm 0cm 0cm, clip}{%
  \importsvg[width=8.5cm]{kurtosis_demo}%
}
    \caption{The normalized skewness and kurtosis of three common distributions as reference to interpret skewness values in terms of the asymmetry of the distribution and kurtosis values in terms of the heavy-tailedness of the distribution.}
\label{fig:skewness_graphic}
\end{figure}

Given the standardized central moment tensors of a distribution, a constrained optimization problem may be solved in order to find the maximal directional skewness and maximal directional excess kurtosis
\begin{equation}
    \max_{\Vert\mathbf{x}\Vert_2=1}T^{i_1,...,i_N}x_{i_1}...x_{i_N}
\end{equation}
This maximal directional skewness or directional excess kurtosis is a measure of the non-Gaussianity of the distribution.
In Fig.~\ref{fig:skewness_graphic}, we show three distributions which serve as a reference for understanding the magnitude of the standardized skewness and standardized excess normalized kurtosis.

\subsection{Monte Carlo-Based Measures}
\label{sec:monte_carlo_based_measures}
We can employ Monte Carlo based techniques to evaluate the measures described in the previous subsection.
These will not be the most efficient, and performing a Monte Carlo subverts the computational motivation of \ac{lincov} analysis.
Instead, these measures will offer a benchmark standard for comparison with the more computationally efficient methods proposed here.
By sampling the initial Gaussian distribution $p(\mathbf{x})$ and mapping each sample $\mathbf{x}_{(i)}$ through the nonlinear function $\mathbf{g}$ we can obtain samples of the distribution of $\mathbf{z}=\mathbf{g}(\mathbf{x})$.
By this approach, we obtain a Dirac mixture approximation of the true mapped density
\begin{align}
  p_{z}(\mathbf{z}) \approx p_{z}^{(\textrm{MC})} (\mathbf{z}) = \frac{1}{N} \sum_{i=1}^{N} \delta(\mathbf{z}-\mathbf{z}_{(i)})
\end{align}
where $ \mathbf{z}_{(i)}=\mathbf{g}(\mathbf{x}_{(i)})$ for $i$ from $1$ to $N$ and $\{\mathbf{x}_{(i)}\}_{i=1}^{N}\sim p(\mathbf{x})$.
Given this reference distribution, the sample mean $\bm{\mu}_{z}^{(\mathrm{MC})}$ and covariance $\mathbf{P}_{z}^{(\mathrm{MC})}$ can be obtained.

The sample distribution can also be used to  statistically analyze the linear propagated moments of
\begin{equation}
    \mathbf{z}^{(1)}=\mathbf{g}(\boldsymbol{\mu}_x)+\mathbf{G}(\mathbf{x}-\boldsymbol{\mu}_x)
\end{equation}
or examine Monte Carlo approximations of the higher-order moments based on the sample skewness $\boldsymbol{\Sigma}^{(\textrm{MC})}$ from~\eqref{eq:skewness} and excess sample kurtosis $\boldsymbol{\kappa}^{(\textrm{MC})}$ from~\eqref{eq:excess_kurtosis} which each rely on the sample mean and covariance.
In particular, the squared Mahalanobis distance of the means, the maximal covariance ratio, the expected squared Mahalanobis distance of the true distribution from the linear covariance distribution can all be calculated by simply using the sample mean and covariance in place of the true mean and covariance as defined in Section~\ref{sec:baseline_monte_carlo_approaches}.
That is,
\begin{align}
  \label{eq:mean-dif_mc}
  \texttt{SMDM}_{\mathbf{g},\mathbf{x}}^{(\mathrm{MC})}
  &=
  (\boldsymbol{\mu}_z^{(\mathrm{MC})}-\boldsymbol{\mu}_z^{(1)})^T(\mathbf{P}^{(1)}_z)^{-1}(\boldsymbol{\mu}_z^{(\mathrm{MC})}-\boldsymbol{\mu}_z^{(1)})\\
  \label{eq:ESMD_mc}
  \texttt{ESMD}_{\mathbf{g},\mathbf{x}}^{(\mathrm{MC})}
  &=
  \operatorname{tr} (  ( \mathbf{P}^{(1)} )^{-1}\mathbf{P}_{z}^{(\mathrm{MC})} ) + \texttt{SMDM}_{\mathbf{g},\mathbf{x}}^{(\mathrm{MC})}
\\
  \label{eq:mcr_mc}
  \texttt{MCR}^{(\mathrm{MC})}
  &=
  \max(1/ \min_{i} \lambda_{i}, \,  \max_{i} \lambda_{i})
\end{align}
where the $\lambda_{i}$ in~\eqref{eq:mcr_mc} are understood to be the eigenvalues of the generalized eigenvalue problem
\begin{align}
  \label{eq:gen_eig_mcr_mc}
  (\mathbf{P}_{z}^{(1)})^{-1}\boldsymbol{\zeta}
  &=
  \lambda (\mathbf{P}_{z}^{(\mathrm{MC})})^{-1}\boldsymbol{\zeta}
\end{align}
Additionally, the maximal directional skewness and kurtosis can be calculated by simply employing the sample mean, covariance, as well as third- and fourth-order central moment tensors in place of the true tensors. On the other hand, the expected squared Mahalanobis distance of the linearization error can be calculated in a Monte Carlo sense using the following procedure:

The linear prediction from any given input data point $\mathbf{x}_{(i)}$ is
\begin{equation}
    \mathbf{z}^{(1)}_{(i)}=\mathbf{z}^{(1)}(\mathbf{x}_{(i)})=\mathbf{g}(\boldsymbol{\mu}_x)+\mathbf{G}(\mathbf{x}_{(i)}-\boldsymbol{\mu}_x)
\end{equation}
and the linearization error for that sample is
\begin{align}
    \delta\mathbf{z}_{(i)}=\mathbf{z}_{(i)}-\mathbf{z}^{(1)}_{(i)}
\end{align}
so the expectation of the squared Mahalanobis distance of the linearization error is approximated by
\begin{equation}
    \texttt{ESMDoLE}_{\mathbf{g},\mathbf{x}}^{(\mathrm{MC})}=\frac{1}{N}\sum_i (\delta\mathbf{z}_{(i)})^T(\mathbf{P}_{z}^{(1)})^{-1}\delta\mathbf{z}_{(i)}
\end{equation}

\subsection{Second-Order and Unscented Approximations of Fidelity Measures}
The Monte Carlo approaches described in the previous subsection provide benchmark measures of fidelity of a given \ac{lincov} approximation.
On the other hand, if the computational resources are available to obtain these Monte Carlo estimates, then the \ac{lincov} results are in effect superseded.
To that end, this subsection develops analytical approximations of the aforementioned measures of fidelity that can be obtained at a fraction of the cost of Monte Carlo approximations.
In pursuit of tractable expressions, two approximation methods are considered: the unscented transform and a second-order Taylor expansion.
The unscented transform is a form of cubature that yields derivative-free approximations of the output mean, covariance, and cross-covariance.
The so-called scaled unscented transform\cite{julier2002ScaledUnscentedTransformation} is summarized below.

Given the input mean $\bm{\mu}_{x}$ and covariance $\mathbf{P}_{x}$, $2n+1$ sigma points and their weights are computed as
\begin{align}
  \footnotesize
\begin{array}{llll}
  \mathcal{X}_{0}=\bm{\mu}_{x} & & w_{0}^{(m)}=\frac{\Lambda}{n+\Lambda} & i=0 \\
  \mathcal{X}_{i}=\bm{\mu}_{x}+\left(\sqrt{(n+\Lambda) \mathbf{P}_{x}}\right)_{i} & i=1, \ldots, n & w_{0}^{(c)}=\frac{\Lambda}{n+\Lambda}+\left(1-\alpha^{2}+\beta\right) & i=0 \\
  \mathcal{X}_{i}=\bm{\mu}_{x}-\left(\sqrt{(n+\Lambda) \mathbf{P}_{x}}\right)_{i-n} & i=n+1, \ldots, 2 n & w_{i}^{(m)}=w_{i}^{(c)}=\frac{1}{2(n+\Lambda)} & i=1, \ldots, 2 n
\end{array}
\end{align}
where the `$(m)$' and `$(c)$' superscripts distinguish the weights used in the mean and covariance approximations, respectively, and where the composite scaling factor is
\begin{align}
\Lambda=\alpha^{2}(n+\kappa)-n
\end{align}
where $\alpha$, $\beta$, and $\kappa$ are free parameters that control the spread of the sigma points and the relative weight of the central sigma point $\mathcal{X}_{0}$; see van de Merwe\cite{merwe2004SigmapointKalmanFilters} for detailed guidance on selecting these parameters.
The output sigma points are then obtained by mapping each sigma point through the full nonlinear function as
\begin{align}
  \mathcal{Z}_{i} = \mathbf{g}(\mathcal{X}_{i}) \qquad i=0,\ldots,2n
\end{align}
The unscented transform approximate mean $\bm{\mu}_{z}^{(\mathrm{UT})}$, covariance $\mathbf{P}_{z}^{(\mathrm{UT})}$ and cross-covariance $\mathbf{P}_{xz}^{(\mathrm{UT})}$ are then given by the weighted sums
\begin{align}
  \label{eq:unscented_mean}
  \bm{\mu}_{z}^{(\mathrm{UT})} & = \sum_{i=0}^{2 n} w_{i}^{(m)} \mathcal{Z}_{i} \\
  \label{eq:unscented_cov}
  \mathbf{P}_{z}^{(\mathrm{UT})} & = \sum_{i=0}^{2 n} w_{i}^{(c)}\left(\mathcal{Z}_{i}-\bm{\mu}_{z}\right)\left(\mathcal{Z}_{i}-\bm{\mu}_{z}\right)^{T} \\
  \label{eq:unscented_cross_cov}
\mathbf{P}_{xz}^{(\mathrm{UT})} & = \sum_{i=0}^{2 n} w_{i}^{(c)}\left(\mathcal{X}_{i}-\mathbf{m}\right)\left(\mathcal{Z}_{i}-\bm{\mu}_{z}\right)^{T}
\end{align}
For Gaussian input distributions, the unscented transform approximation is exact for expectations of polynomials up to degree three and thus is generally more accurate than \ac{lincov} approximations.

The second approximation method considered is based on a Taylor series expansion about the input mean:
\begin{equation}
    \label{eq:2nd_order_mean}
    \mathbf{g}(\boldsymbol{\mu}_x+\mathbf{x})\approx\mathbf{g}(\boldsymbol{\mu}_x)+\mathbf{G}\delta\mathbf{x}+\frac{1}{2}\mathbf{G}^{(2)}\delta\mathbf{x}^2
\end{equation}
where
\begin{equation}
    \mathbf{G}^{(2)}=\frac{\partial^2 \mathbf{g}}{\partial \mathbf{x}^2}\Big\vert_{\boldsymbol{\mu}_x}
\end{equation}
and the double contraction in shorthand is explicitly in Einstein notation
\begin{equation}
    (\mathbf{G}^{(2)}\delta\mathbf{x}^2)^i=(\mathbf{G}^{(2)})^i_{j,k}\delta\mathbf{x}^j\delta\mathbf{x}^k
\end{equation}

\textbf{Expected Squared Mahalanobis Distance of Second-Order Approximation of Linearization Error}

Consider a deviation from the input mean:
\begin{equation}
    \delta\mathbf{x}=\mathbf{x}-\boldsymbol{\mu}_x
\end{equation}
The error in the approximation of $\mathbf{g}(\mathbf{x})$ with $\mathbf{g}(\boldsymbol{\mu}_x)+\mathbf{G}\delta\mathbf{x}$ will be dominated by a second-order contribution assuming $\delta\mathbf{x}$ is sufficiently small.
This second-order contribution is given by the second term in the Taylor series~\eqref{eq:2nd_order_mean}.

Thus, the random vector $\frac{1}{2}\mathbf{G}^{(2)}\delta\mathbf{x}^2$ dominates the linearization error and its magnitude serves as a measure of nonlinearity at the deviation $\delta \mathbf{x}$.
The expected value of this error expressed as a squared Mahalanobis distance is
\begin{align}
\label{eq:expected_squared_mahalanobis_distance_linearization_error_second_order}
\texttt{ESMDoLE}_{\mathbf{g},\mathbf{x}}^{(2)}
=
  \mathrm{E}
  \left[
    \frac{1}{4}
    \left(
    \mathbf{G}^{(2)}\delta\mathbf{x}^2
   \right)^{T}
   (\mathbf{P}_{z}^{(1)})^{-1}
    \left(
    \mathbf{G}^{(2)}\delta\mathbf{x}^2
   \right)
  \right]
\end{align}
where the `$(2)$' superscript in $\texttt{ESMDoLE}_{\mathbf{g},\mathbf{x}}^{(2)}$ indicates that this term is a second-order approximation of~\eqref{eq:expected_squared_mahalanobis_distance_linearization_error}.
The expectand in~\eqref{eq:expected_squared_mahalanobis_distance_linearization_error_second_order} is written explicitly in Einstein notation as
\begin{equation}
  \label{eq:second_order_squared_md_linearization_error}
    \frac{1}{4}((\mathbf{P}_z^{(1)})^{-1})_{i_1,i_2}(\mathbf{G}^{(2)})^{i_1}_{j_1,k_1}\delta\mathbf{x}^{j_1}\delta\mathbf{x}^{k_1}(\mathbf{G}^{(2)})^{i_2}_{j_2,k_2}\delta\mathbf{x}^{j_2}\delta\mathbf{x}^{k_2}
\end{equation}
Unlike the Monte Carlo based approximation, the second-order approximation admits a closed-form expression
\begin{equation}
\label{eq:expected_squared_mahalanobis_distance_linearization_error_second_order_kurtosis}
        \texttt{ESMDoLE}_{\mathbf{g},\mathbf{x}}^{(2)}=\frac{1}{4}((\mathbf{P}_z^{(1)})^{-1})_{i_1,i_2}(\mathbf{G}^{(2)})^{i_1}_{j_1,j_2}(\mathbf{G}^{(2)})^{i_2}_{j_3,j_4}(\mathbf{K}_x)^{j_1,j_2,j_3,j_4}
\end{equation}
where $\mathbf{K}_x$ is the fourth-order central moment tensor for the random variable $\mathbf{x}$.
Furthermore, if $\mathbf{x}$ is Gaussian distributed, the four-order central moment can be obtained in terms of the covariance matrix as
\begin{align}
    (\mathbf{K}_x)^{j_1,j_2,j_3,j_4} &= \mathrm{E}[\delta\mathbf{x}^{j_1}\delta\mathbf{x}^{j_2}\delta\mathbf{x}^{j_3}\delta\mathbf{x}^{j_4}]\\
    &= \operatorname{sym}\left(3(\mathbf{P}_x)^{j_1,j_2}(\mathbf{P}_x)^{j_3,j_4}\right)
\end{align}
Note that we do not present an unscented transform version of this expected squared Mahalanobis distance of linearization error because the third-order accurate unscented transform cannot accurately quantify this expectation which is inherently a fourth-order quantity in the deviation from the initial mean.

\textbf{Mahalanobis Distance of the First- and Second-Order Means}

The second-order mean is calculated as the expectation of~\eqref{eq:2nd_order_mean} as
\begin{equation}
    \label{eq:mean_second_order}
    \left(\boldsymbol{\mu}^{(2)}_z\right)^i=\mathbf{g}(\boldsymbol{\mu}_x)^i+(\delta\boldsymbol{\mu}_z^{(2)})^i
\end{equation}
where
\begin{equation}
\label{eq:dmu2}
(\delta\boldsymbol{\mu}_z^{(2)})^i=\frac{1}{2}\left(\mathbf{G}^{(2)}\right)^i_{j,k}(\mathbf{P}_x)^{j,k}
\end{equation}
Similarly, the difference between the unscented approximation of the mean and the first-order approximation can be obtained as
\begin{align}
  \delta \bm{\mu}_{z}^{(\mathrm{UT})} = \bm{\mu}_{z}^{(\mathrm{UT})} - \mathbf{g}(\bm{\mu}_{x})
\end{align}

The second-order approximation~\eqref{eq:mean_second_order} and unscented approximation of the mean can be compared to the first-order approximation of the mean using a linear covariance Mahalanobis distance as described in~\eqref{eq:mean-dif}, giving the second-order and unscented variants of the squared Mahalanobis distance of the means:
\begin{align}
    \texttt{SMDM}_{\mathbf{g},\mathbf{x}}^{(2)}
    &=
    (\delta\boldsymbol{\mu}_{z}^{(2)})^T (\mathbf{P}_{z}^{(1)})^{-1} (\delta\boldsymbol{\mu}_{z}^{(2)})\\
    \texttt{SMDM}_{\mathbf{g},\mathbf{x}}^{(\mathrm{UT})}
    &=
    (\delta\boldsymbol{\mu}_{z}^{(\mathrm{UT})})^T (\mathbf{P}_{z}^{(1)})^{-1} (\delta\boldsymbol{\mu}_{z}^{(\mathrm{UT})})
\end{align}

\textbf{Maximal Ratio Between First- and Second-Order Covariance Ellipsoids}

The covariance of the random variable $\mathbf{z}$ as determined up to a linear approximation of the function $\mathbf{g}$ is given by~\eqref{eq:lin_cov}.
An approximation of the covariance matrix up to second-order is given by \cite{park2006nonlinear}
\begin{equation}
    \label{eq:2nd_order_cov}
    (\mathbf{P}_{z}^{(2)})^{i_1,i_2}=(\mathbf{G}\mathbf{P}_x\mathbf{G}^T)^{i_1,i_2}+\frac{1}{4}(\mathbf{G}^{(2)})^{i_1}_{j_1,j_2}(\mathbf{G}^{(2)})^{i_2}_{j_3,j_4}(\mathbf{K}_x)^{j_1,j_2,j_3,j_4}-(\delta\boldsymbol{\mu}_{z}^{(2)})^{i_1}(\delta\boldsymbol{\mu}_{z}^{(2)})^{i_2}
\end{equation}
We may examine the ratio of equal likelihood covariance ellipsoids along a given direction using the generalized Rayleigh quotient
\begin{equation}
\frac{\boldsymbol{\zeta}^T(\mathbf{P}_z^{(1)})^{-1}\boldsymbol{\zeta}}{\boldsymbol{\zeta}^T(\mathbf{P}_z^{(2)})^{-1}\boldsymbol{\zeta}}
\end{equation}
The generalized Rayleigh quotient is minimized or maximized by solving the generalized eigenvalue problem with
\begin{equation}
\label{eq:mcr_gen_eig_second_order}
    (\mathbf{P}_z^{(1)})^{-1}\boldsymbol{\zeta}=\lambda (\mathbf{P}_z^{(2)})^{-1}\boldsymbol{\zeta}
\end{equation}
and finding the minimum and maximum eigenvalue which correspond to these minimum and maximum ratios.
That is, the \ac{mcr} measure based on the second-order approximation is
\begin{align}
    \texttt{MCR}^{(2)} = \max(1/ \min_{i} \lambda_{i}, \,  \max_{i} \lambda_{i})
\end{align}
where the eigenvalues $\lambda_{i}$ are understood to be those of~\eqref{eq:mcr_gen_eig_second_order}.
As discussed in Section~\ref{sec:baseline_monte_carlo_approaches}, this measure be interpreted as giving the maximum and minimum $\mathbf{P}^{(1)}$-induced Mahalanobis distance of the 1-sigma ellipsoid based on $\mathbf{P}^{(2)}$.

Note that the eigenvalues of the pair $\left((\mathbf{P}_z^{(1)})^{-1}, (\mathbf{P}_z^{(2)})^{-1}\right)$ are the same as the eigenvalues of the pair $\left(\mathbf{P}_z^{(2)}, \mathbf{P}_z^{(1)}\right)$.
Thus, the maximum or  minimum-reciprocal ratios of the precisions are the same as the minimum-reciprocal or maximum ratios of the variances along any direction, respectively.
Computing the generalized eigenvalues of the covariance matrices is more numerically stable and efficient than computing inverses and then computing the generalized eigenvalues of these.

We may repeat this same analysis as in the previous section, but rather than computing the second-order approximation of the covariance using the higher-order partial derivative approach, the higher-order approximation of the covariance may be computed using the unscented transform to find $\mathbf{P}_z^{(\mathrm{UT})}$.
From here, the generalized eigenvalue analysis with the inverses of $\mathbf{P}_z^{(\mathrm{UT})}$ and $\mathbf{P}_z^{(1)}$ can be conducted.
This can be interpreted as measuring the difference between the uncertainty propagation of an unscented Kalman filter and an extended Kalman filter.

\textbf{Expected Squared Mahalanobis Distance Between Linear and Higher-Order Mappings}

Similar to the analysis done with the expected squared Mahalanobis distance of the true distribution, the expected squared Mahalanobis distance can be approximated by comparing the linear propagation of the mean and covariance with the second-order propagation of the mean and covariance.
Given these expressions, the expected squared Mahalanobis distance can be found using~\eqref{eq:ESMD} replacing the true mean and covariance with their second-order and unscented approximations to give
\begin{align}
\label{eq:ESMD_second_order}
\texttt{ESMD}_{\mathbf{g},\mathbf{x}}^{(2)}
&= \operatorname{tr} (  ( \mathbf{P}_{z}^{(1)} )^{-1}\mathbf{P}_{z}^{(2)}  ) + (\delta\boldsymbol{\mu}_{z}^{(2)})^T (\mathbf{P}_{z}^{(1)})^{-1} (\delta\boldsymbol{\mu}_{z}^{(2)})\\
\label{eq:ESMD_unscented}
\texttt{ESMD}_{\mathbf{g},\mathbf{x}}^{(\mathrm{UT})}
&= \operatorname{tr} (  ( \mathbf{P}_{z}^{(1)} )^{-1}\mathbf{P}_{z}^{(\mathrm{UT})}  ) + (\delta\boldsymbol{\mu}_{z}^{(\mathrm{UT})})^T (\mathbf{P}_{z}^{(1)})^{-1} (\delta\boldsymbol{\mu}_{z}^{(\mathrm{UT})})
\end{align}
where the second-order covariance $\mathbf{P}_{z}^{(2)}$ is obtained using~\eqref{eq:2nd_order_cov} and $\delta\boldsymbol{\mu}_{z}^{(2)}$ is obtained with~\eqref{eq:dmu2}.

\textbf{Maximal Normalized Skewness and Kurtosis Analysis}

Computing a second- or higher-order approximation of the skewness and kurtosis of~$\mathbf{z}$ can be accomplished using partial derivative tensors \cite{majji2008jth,majji2010perturbation}.
Consider the central third-order moment tensor of $\mathbf{z}$.
Employing the second-order approximation of the nonlinear function $\mathbf{g}$, we have that the corresponding central third-order moment tensor is
\begin{align}
    (\mathbf{S}^{(2)}_z)^{i,j,k}=&\frac{1}{2}(\mathbf{G}^i_{l_1}\mathbf{G}^j_{l_2}(\mathbf{G}^{(2)})^k_{l_3,l_4}+\mathbf{G}^i_{l_1}\mathbf{G}^k_{l_2}(\mathbf{G}^{(2)})^j_{l_3,l_4}+\mathbf{G}^j_{l_1}\mathbf{G}^k_{l_2}(\mathbf{G}^{(2)})^i_{l_3,l_4})(\mathbf{K}_x)^{l_1,l_2,l_3,l_4}\nonumber\\
    &+\frac{1}{8}\mathbf{G}^i_{l_1,l_2}\mathbf{G}^j_{l_3,l_4}\mathbf{G}^k_{l_5,l_6}(\mathbf{K}^{[6]}_x)^{l_1,...,l_6} \nonumber\\
    &-
    3 \operatorname{sym}((\mathbf{P}_z^{(2)})^{i,j}(\delta\boldsymbol{\mu}_z^{(2)})^k)+2(\delta\boldsymbol{\mu}_z^{(2)})^i(\delta\boldsymbol{\mu}_z^{(2)})^j(\delta\boldsymbol{\mu}_z^{(2)})^k\nonumber
\end{align}
because the third-order central moments of $\mathbf{x}$ are all zero.
The notation $\mathbf{K}^{[2M]}$ denotes the even $2M$-th order central moment tensor (generalizing the 4th order moment tensor $\mathbf{K}$).
The even higher-order central moments of a Gaussian distribution are given in terms of the covariance as
\begin{align}
    (\mathbf{K}^{[2M]})^{i_1,..., i_{2M}}&=\frac{(2M)!}{2^M(M)!}\operatorname{sym}(\mathbf{P}^{i_1,i_2}...\mathbf{P}^{i_{2M-1},i_{2M}})
\end{align}
where the constant out front is the number of ways to partition a set of $2M$ objects into pairs when order does not matter.

The central fourth-order moment tensor up to second-order is given by
\begin{align}
    (\mathbf{K}_z^{(2)})^{i,j,k,l}&=\mathbf{G}^i_{q_1}\mathbf{G}^j_{q_2}\mathbf{G}^k_{q_3}\mathbf{G}^l_{q_4}(\mathbf{K}_x)^{q_1,q_2,q_3,q_4}\\
    &+ \frac{1}{4}\binom{4}{2}\operatorname{sym}((\mathbf{G}^{(2)})^i_{q_1,q_2}(\mathbf{G}^{(2)})^j_{q_3,q_4}\mathbf{G}^k_{q_5}\mathbf{G}^l_{q_6}(\mathbf{K}_x^{[6]})^{q_1,...,q_6})\nonumber\\
    &+ \frac{1}{16}(\mathbf{G}^{(2)})^i_{q_1,q_2}(\mathbf{G}^{(2)})^j_{q_3,q_4}(\mathbf{G}^{(2)})^k_{q_5,q_6}(\mathbf{G}^{(2)})^l_{q_7,q_8}(\mathbf{K}_x^{[8]})^{q_1,...,q_8}\nonumber\\
    &-4\operatorname{sym}((\mathbf{S}^{(2)}_z)^{i,j,k})(\delta\boldsymbol{\mu}_z^{(2)})^l)\nonumber\\
    &+6\operatorname{sym}((\mathbf{P}_z^{(2)})^{i,j}(\delta\boldsymbol{\mu}_z^{(2)})^k(\delta\boldsymbol{\mu}_z^{(2)})^l)-3(\delta\boldsymbol{\mu}_z^{(2)})^i(\delta\boldsymbol{\mu}_z^{(2)})^j(\delta\boldsymbol{\mu}_z^{(2)})^k(\delta\boldsymbol{\mu}_z^{(2)})^l\nonumber
\end{align}
where the binomial coefficient of the second term is ``4 choose 2" (which is equal to 6).
The symmetrization of a tensor is given in terms of the average over all permutations of the indices as
\begin{equation}
    \operatorname{sym}(\mathbf{T}^{i_1...i_N})=\frac{1}{N!}\sum_{\sigma\in S_{N} }\mathbf{T}^{\sigma(i_1,...,i_N)}
\end{equation}
If there are underlying symmetries, the sum need not be computed over all permutations so long as the normalization factor is appropriately adjusted.
Another technique for computing the higher-order moment tensors not pursued here involves cubature based methods such as the conjugate unscented transform \cite{adurthi2012conjugate}.

Once the central moment tensors are known, the normalized central moment tensors can be calculated by applying the whitening transformation to each index of the central-moment tensor where the whitening transformation is given by the inverse matrix square root of the second-order approximation of the covariance.

\section{Optimization-Based Fidelity Measures}
\label{sec:optimization_based_fidelity_measures}
In previous work, constrained optimization-based methods were employed to determine the need for and direction of Gaussian mixture splitting in the context of nonlinear uncertainty propagation \cite{kulik2025NonlinearityUncertaintySplitting}.
We summarize a few of the methods introduced in that work that provide the most easily interpretable metrics associated with the performance of \ac{lincov} approximation.

\subsection{Maximal Mahalanobis Distance of 1-sigma Linearization Error}

\begin{figure}[htbp]
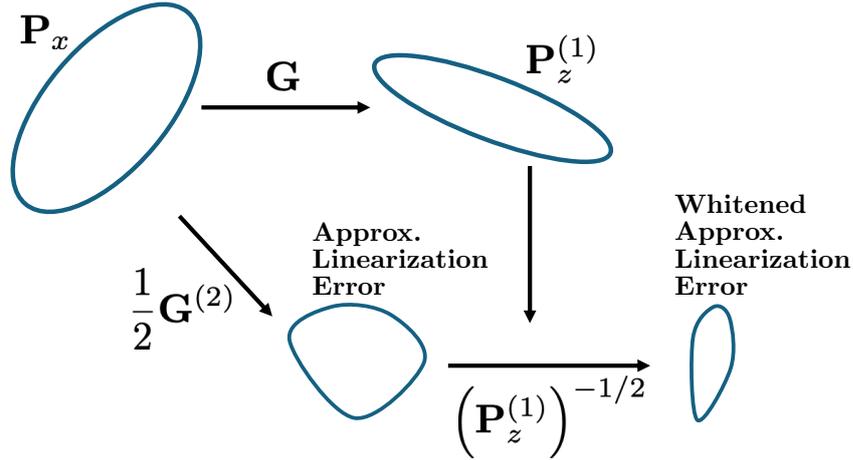

    \centering
    \importsvg[width=.8\linewidth]{whitened_nonlinearity_diagram}
    \caption{A schematic describing the quantity being maximized in computing \ac{wussos}.}
    \label{fig:wussos_diagram}
\end{figure}

The \ac{wussos} heuristic
\begin{equation}
  \label{eq:hybrid_second_order_stretching}
    \max_{\mathbf{x}^T\mathbf{P}_x^{-1}\mathbf{x}=1}\Vert \mathbf{G}^{(2)}\mathbf{x}^2\Vert_{\mathbf{P}_z^{-1}}
\end{equation}
 combines nonlinearity, uncertainty, and output whitening to characterize the effectiveness of linear uncertainty propagation
 \cite{kulik2025NonlinearityUncertaintySplitting}.
 \Ac{wussos} gives the maximum Mahalanobis distance (induced by the linearly propagated covariance) of the second-order approximation of the linearization error for any initial point on a manifold of equal probability density in the input space.
Fig.~\ref{fig:wussos_diagram} depicts the transformations involved in \ac{wussos}, the constraint manifold, and the whitened manifold over which the stretching distance from the origin is maximized.

The \ac{wussos} optimization problem can be rephrased in a manner computable as a maximal Z-eigenvector with shifted symmetric higher-order power iteration as
\begin{equation}
    \max_{\mathbf{y}^T\mathbf{y}=1}  (\mathbf{G}^{(2)}(\mathbf{P}_x^{1/2}\mathbf{y})^2)^T \mathbf{P}_z^{-1} (\mathbf{G}^{(2)}(\mathbf{P}_x^{1/2}\mathbf{y})^2)
    \label{eq:whitened_second_order}
\end{equation}
The relevant fourth-order tensor whose square-rooted maximum Z-eigenvalue gives \ac{wussos} can be written as
\begin{equation}
    \mathbf{W}_{i_1,i_2,i_3,i_4}=\left(\mathbf{P}_z^{-1} \right)_{j_1,j_2}\left(\mathbf{G}^{(2)}\right)^{j_1}_{k_1,k_2}\left(\mathbf{G}^{(2)}\right)^{j_2}_{k_3,k_4}\left(\mathbf{P}_x^{1/2}\right)^{k_1}_{i_1}...\left(\mathbf{P}_x^{1/2}\right)^{k_4}_{i_4}
\end{equation}
\subsection{Maximal 1-sigma Frobenius Norm Error of the Linearization}
Tuggle and Zanetti \cite{tuggle2018automated,tuggle2020model} first introduced a Gaussian mixture splitting method that employed nonlinear function properties and uncertainty scaling.
In that work, the splitting direction is chosen according to the optimization problem
\begin{equation}
    \delta\hat{\mathbf{x}}^*\sim \argmax_{\delta\mathbf{x}^T\mathbf{P}_x^{-1}\delta\mathbf{x}=1}\Vert \mathbf{G}^{(2)}\delta\mathbf{x}\Vert_{F}
\end{equation}
where the subscript $F$ denotes the Frobenius norm and $\mathbf{G}^{(2)}\delta\mathbf{x}$ denotes the matrix with components
\begin{equation}
    \left(\mathbf{G}^{(2)}\delta\mathbf{x}\right)^{i}_{j}=(\mathbf{G}^{(2)})^i_{j,k} \delta x^k
\end{equation}
which describes a linear approximation of how the Jacobian of $\mathbf{g}$ changes as it is evaluated at a point $\delta\mathbf{x}$ away from the current point it is being evaluated at.
This splitting direction and the associated metric can be calculated using a generalized eigenvalue problem with the two matrices $(\bar{\mathbf{G}}^{(2)},\mathbf{P}_x^{-1})$
or with the singular value decomposition of $\bar{\mathbf{G}}^{(2)}\mathbf{P}_x^{1/2}$ where the $mn$ by $n$ matricization of the tensor $\mathbf{G}^{(2)}$ is
\begin{equation}
    (\bar{\mathbf{G}}^{(2)})^{ni+j}_k=(\mathbf{G}^{(2)})^i_{j,k}
\end{equation}

In our previous work\cite{kulik2025NonlinearityUncertaintySplitting}, we established a scale invariance for this metric and selection criterion by transforming the linear transformation given by the matrix $\mathbf{G}^{(2)}\delta\mathbf{x}$ so that it maps from a whitened input space to a whitened output space.
For completeness, we summarize this process here.
The input whitening transformation is given by
\begin{equation}
    \delta\mathbf{x}'=\mathbf{P}_x^{-1/2}\delta\mathbf{x}
\end{equation}
and the whitening transformation for the output space is given by
\begin{equation}
    \mathbf{z}'=(\mathbf{P}_z^{(1)})^{-1/2}\mathbf{z}
\end{equation}
A generic linear transformation $\mathbf{A}$ from the original input space to the original output space can be transformed to a linear transformation from the whitened input space to the whitened output space according to
\begin{equation}
    \mathbf{A}'= (\mathbf{P}_{z}^{(1)})^{-1/2} \mathbf{A} \mathbf{P}_x^{1/2}
\end{equation}
If we apply this transformation to the linear transformation given by the matrix $\mathbf{G}^{(2)}\delta\mathbf{x}$, we obtain a linear transformation %
given by the matrix
\begin{equation}
    (\mathbf{P}_{z}^{(1)})^{-1/2}(\mathbf{G}^{(2)}\delta\mathbf{x})\mathbf{P}_x^{1/2}
\end{equation}
whose squared Frobenius norm will characterize the change in the linear approximation of the nonlinear function $\mathbf{g}$ (normalized to map between whitened input and output spaces) at a step $\delta\mathbf{x}$ away from the current point of linearization.

\begin{figure}
    \centering
    \adjustbox{trim=0cm 0cm 11cm 0cm, clip}{%
    \importsvg[width=1.4\linewidth]{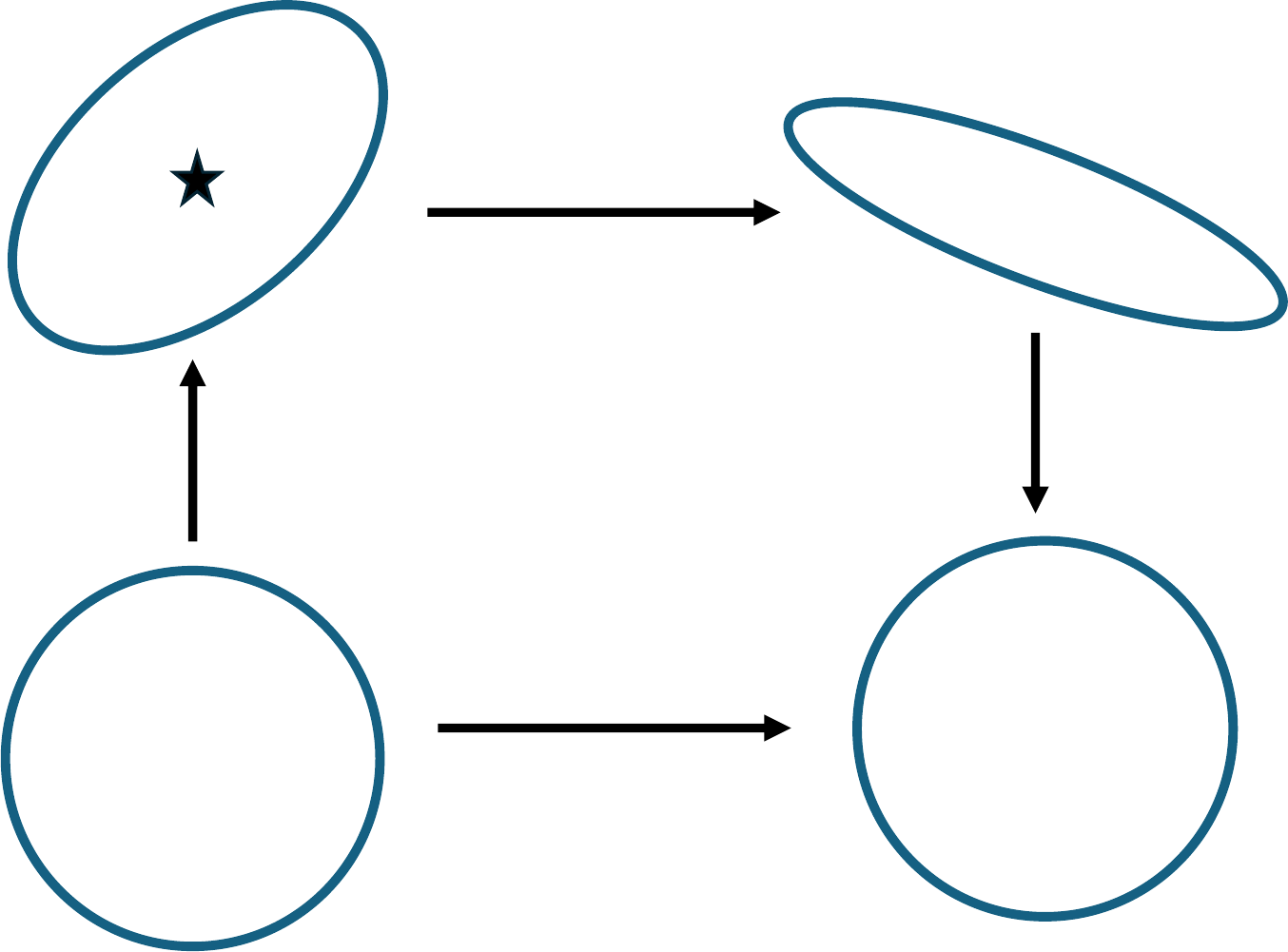}
    }
    \caption{A schematic describing the reference transformation for comparison with the transformations being optimized in W-US-SOLC.}
    \label{fig:wussolc_diagram1}
\end{figure}

In order to assess whether the change in this whitened linearization is significant we should compare against the squared Frobenius norm of the original linearization between whitened spaces.
If the original linear transformation is $\mathbf{A}=\mathbf{G}$ and both $\mathbf{P}_x,\mathbf{P}_z^{(1)}$ are nonsingular, then the corresponding linear transformation between whitened spaces $\mathbf{G}'$ is a linear transformation from the unit sphere in $n$ dimensions (the dimension of the domain of $\mathbf{g}$) to the unit sphere in $m$ dimensions (the dimension of the codomain of $\mathbf{g}$).
Thus, $\mathbf{G}'$ is an orthogonal matrix.
The transformation $\mathbf{G}'$ is illustrated in the bottom line of Fig.~\ref{fig:wussolc_diagram1}.
Note that since the eigenvalues of an orthogonal matrix only take values $-1,0$, and $1$, the rank of the matrix gives the number of nonzero eigenvalues, and the Frobenius norm of $\mathbf{G}'$ is given by the sum of squares of the eigenvalues
\begin{equation}
\Vert\mathbf{G}'\Vert_F^2=\min(n,m)
\end{equation}
As such, the squared Frobenius norm of the whitened change in the linearization given by $\mathbf{G}^{(2)}\delta\mathbf{x}$ can naturally be compared to $\min(n,m)$. If this value is small relative to $\min(n,m)$, then the linearization at a point $\delta\mathbf{x}$ away from the mean would propagate the initial covariance ellipsoid to a final covariance ellipsoid that is very similar.

The resulting \ac{wussolc} metric
\begin{equation}
    \max_{\delta\mathbf{x}^T\mathbf{P}_x^{-1}\delta\mathbf{x}=1 }\Vert (\mathbf{P}_{z}^{(1)})^{-1/2}(\mathbf{G}^{(2)}\delta\mathbf{x})\mathbf{P}_x^{1/2}\Vert_{F}^2 = \max_{\mathbf{y}^T\mathbf{y}=1}\Vert (\mathbf{P}_{z}^{(1)})^{-1/2}(\mathbf{G}^{(2)}(\mathbf{P}^{1/2}_x\mathbf{y})))\mathbf{P}_x^{1/2}\Vert_{F}^2
\end{equation}
can be compared to $\min(n,m)$ as a measure of the maximal normalized change in the linearization over the 1-$\sigma$ covariance ellipsoid.
Note that this metric corresponds to the induced (Frobenius, 2)-norm of the same tensor of which \ac{wussos} is the induced 2-norm.
See previous work on tensor norms for details on the relationship between the induced 2-norm of a tensor and the induced (Frobenius, 2)-norm of a tensor and the inequality between the two \cite{kulik2024applications}. In order to compute \ac{wussolc} most simply, compute the tensor
\begin{equation}
    (\mathbf{G}_w^{(2)})^i_{j,k}=((\mathbf{P}_{z}^{(1)})^{-1/2})^i_l(\mathbf{G}^{(2)})^l_{p,q}(\mathbf{P}_x^{1/2})^p_j(\mathbf{P}_x^{1/2})^q_k
\end{equation}
and its $mn$ by $n$ matricization
\begin{equation}
    (\bar{\mathbf{G}}^{(2)}_w)^{ni+j}_k=(\mathbf{G}^{(2)}_w)^i_{j,k}
\end{equation}
from which we can write the \ac{wussolc} metric as
\begin{equation}
    \max_{\mathbf{y}^T\mathbf{y}=1}\Vert \bar{\mathbf{G}}^{(2)}_w\mathbf{y}\Vert_{2}^2
\end{equation}
which is simply the maximal singular value of the matrix $\bar{\mathbf{G}}^{(2)}_w$.
\subsection{Maximal Mahalanobis Distance from Statistical and Deterministic Linearization Error}
The previous two methods rely on computation of higher-order derivative tensors which can be costly and difficult to implement\footnote{Python implementations of \ac{wussos}, \ac{wussolc}, and other measures are publicly available in the PyEst library: \url{https://github.com/scope-lab/pyest}} for the uninitiated working in this framework.
Unscented transform-based methods offer an advantage in that the evaluation of the nonlinear transformation $\mathbf{g}(\mathbf{x})$ at the sigma points (or more generally, regression points) can be compared to the linearized system to assess nonlinearity without having to directly calculate second-order derivatives \cite{kulik2025NonlinearityUncertaintySplitting}.

Consider the statistical linearization
\begin{align}
  \mathbf{z} = \mathbf{g}(\mathbf{x}) \approx \mathbf{G}^{(\textrm{SL})} \mathbf{x} + \mathbf{b}
\end{align}
where the `$(\mathrm{SL})$' superscript indicates a statistical linearization term.
The matrix $\mathbf{G}^{(\textrm{SL})}$ and vector $\mathbf{b}$ that minimize the mean squared error for this affine model over some sampled points are
\begin{align}
  \label{eq:statistical_linearization_matrix}
  \mathbf{G}^{(\textrm{SL})}
  =
  \left(\mathbf{P}_{x z}^{(\mathrm{UT})}\right)^{T}\left(\mathbf{P}_x\right)^{-1} \text { and } \mathbf{b}=\boldsymbol{\mu}_z^{(\mathrm{UT})}-\mathbf{G}^{(\textrm{SL})} \boldsymbol{\mu}_x
\end{align}
where the output mean $\boldsymbol{\mu}_{z}^{(\mathrm{UT})}$ and cross-covariance $\mathbf{P}_{xz}^{(\mathrm{UT})}$ are readily obtained via the unscented transform in~\eqref{eq:unscented_mean} and~\eqref{eq:unscented_cross_cov}, respectively, or other sigma point transformation  \cite{huber2011AdaptiveGaussianMixture}.
Given the true Jacobian $\mathbf{G}$ of the nonlinear function $\mathbf{g}$ evaluated at the mean and the statistical linearization $\mathbf{G}^{(\textrm{SL})}$, another interesting quantity is their difference.
The \ac{sadl} calculates the induced 2-norm of the difference $\mathbf{G}^{(\textrm{SL})}-\mathbf{G}$.

We can extend the method to take into account initial uncertainty and rescale the final coordinates according to the final covariance.
The resulting metric is then the matrix 2-norm
\begin{equation}
  \Vert (\mathbf{P}_{z}^{(1)})^{-1/2}(\mathbf{G}^{(\mathrm{SL})}-\mathbf{G})\mathbf{P}^{1/2}_x\Vert_{2}
    \label{eq:stat_lin_dif_full}
\end{equation}
The covariance $\mathbf{P}_z$ above can either be the covariance obtained from the sigma point method or the covariance given by propagating the original covariance using the deterministic linearization.
In this work, we will employ the covariance arising from the deterministic linearization.
This metric produces the largest error in the statistical and deterministic linearizations as applied to vectors on the original 1-sigma covariance ellipsoid in terms of the $\mathbf{P}_z$-Mahalanobis distance.
This method will be referred to as the \ac{wussadl} method. This metric is connected to an optimization method for maximal nonlinearity direction analysis presented in Kulik and LeGrand\cite{kulik2025NonlinearityUncertaintySplitting}.

\section{Applications}
\label{sec:applications}
In this section, we test the aforementioned methods in an analysis involving the proposed NASA Gateway \ac{nrho}.
An initial Gaussian state uncertainty is assumed, and the distribution is propagated using linear covariance techniques over the course of an orbit.
Over that period, we compare the  proposed analytic metrics with Monte Carlo based metrics to assess their accuracy and ultimate ability to identify divergent LinCov results.

\subsection{Three-Body Motion}
\label{sec:three_body_motion}
For simplicity, we adopt the circular three-body problem assumptions.
The corresponding equations of motion expressed in the synodic frame are
\begin{align}
    \frac{\mathrm{d}}{\mathrm{d}t}\mathbf{x}&=\mathbf{F(x)}\\
    \mathbf{F(x)}&=\begin{bmatrix} \dot{x}& \dot{y}& \dot{z}& 2\dot{y}+\frac{\partial \overline{U}}{\partial x}& -2\dot{x}+\frac{\partial \overline{U}}{\partial y}& \frac{\partial \overline{U}}{\partial z}\end{bmatrix}^T
    \label{eqn:cr3bp}
\end{align}
where $\overline{U}(x,y,z)=\dfrac{1-\mu^*}{||\mathbf{r}_1||}+\dfrac{\mu^*}{||\mathbf{r}_2||}+\dfrac{x^2+y^2}{2}$ is the effective potential and the mass ratio is defined as $\mu^*=\dfrac{m_2}{m_1+m_2}$ for the two primary bodies with masses $m_1,m_2$ respectively.
We adopt the convention that the primary body with greater mass is assigned the index $1$ so that $m_1\geq m_2$.
Both masses lie on the $x$-axis at $[-\mu^*,0,0]$ and $[1-\mu^*, 0, 0]$ with respect to their common barycenter at the origin.
The positions of the satellite of interest with respect to the primary and secondary bodies are denoted by $\mathbf{r}_1$ and $\mathbf{r}_2$ respectively  \cite{koon2000dynamical}.

\begin{figure}[htpb]
  \centering
  
  \ifimporttikz
    \tikzsetnextfilename{cislunar_trajectory}
    \import{Figures/}{cislunar_trajectory.tikz}
  \else
    \includegraphics{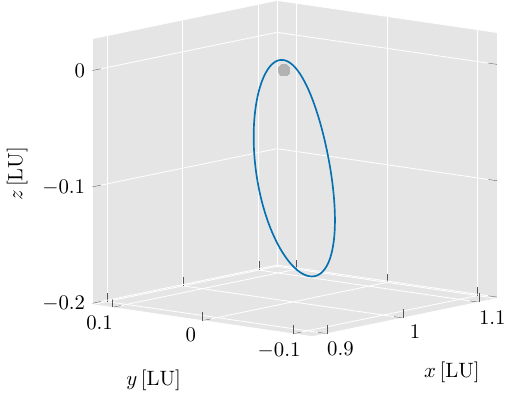}
  \fi

  \caption{Full NRHO considered in three-body uncertainty propagation application.}%
  \label{fig:nrho}
\end{figure}

The reference orbit used in the following sections is a 9:2 resonant Southern $L_{2}$ halo orbit like that proposed for the NASA Gateway \cite{NationalAA2019} and shown in Figure~\ref{fig:nrho}. These initial conditions were obtained as a canonical unit conversion of the initial conditions used in the QIST model of Gateway \cite{cunningham2023interpolated}.
The initial conditions (coinciding with apolune) and mass parameter used for the orbit are
\begin{gather*}
    \mu=1.0/(81.30059 + 1.0), \quad x_0=1.022022,\\z_0 = -0.182097, \quad \dot{y}_0 = -0.103256
\end{gather*}
in nondimensional units with other initial coordinates equal to zero.

The period of the orbit is approximately $1.511111 \, [\textrm{TU}]$ where
\begin{equation}
    2 \pi [\textrm{TU}] = 2.361 \cdot 10^{6} \, [\textrm{sec}]
\end{equation}

In this example, the nonlinear function $\mathbf{g}$ in question is the flow of the circular restricted three-body dynamics $\varphi_t$ for some fixed value of the time-of-flight $t$.
The flow map associated with the dynamical system in~\eqref{eqn:cr3bp} is defined such that
\begin{equation}
\left(\frac{\mathrm{d}}{\mathrm{d}\tau}\varphi_\tau(\mathbf{x}_0)\right)\bigg\rvert_{\tau=t}=\mathbf{F}(\varphi_t(\mathbf{x}_0)), \quad \varphi_0(\mathbf{x}_0)=\mathbf{x}_0
\end{equation}
The Jacobian and the second-order partial derivative tensor of the flow map around some reference trajectory are the state transition matrix and the second-order state transition tensor, respectively.
These quantities are obtained by integrating the variational equations \cite{park2006nonlinear}, or by employing techniques from differential algebra \cite{rasotto2016differential}.
While repeated computation of state transition tensors can be costly, techniques exist for precomputing them offline along a known reference trajectory and then efficiently interpolating the state transition tensors online\cite{cunningham2023interpolated,kulik2023state,cunningham2024spice}.
Other efficient approaches are available that leverage the potential low-rank qualities of the state transition tensors \cite{boone2023directional,boone2024efficient,zhou2024time}.

\begin{figure}
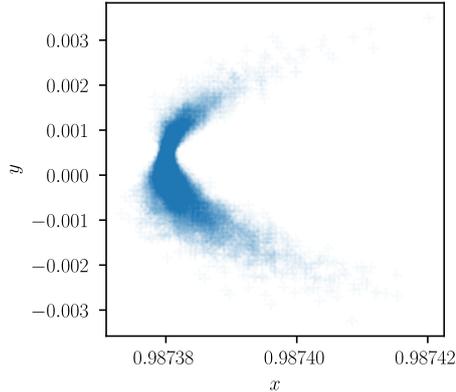

    \centering
\adjustbox{trim=0cm 0cm 0cm 0cm, clip}{%
  \importsvg[width=6cm]{cislunartruth_scatter_0_1}%
}
\caption{Scatter plot of Monte Carlo samples at reference orbit perilune in the x-y plane.}
\label{fig:scatter}
\end{figure}

In this example, we employ an initial Gaussian distribution with mean equal to the Gateway-like initial conditions and covariance given by
\begin{equation}
    \mathbf{P}_x=10^{-8}\mathrm{diag}([1,0,1,0,0,0])+10^{-10}\mathbf{I}_6
\end{equation}
in nondimensional canonical units.
The 1-sigma distances are on the order of $40 \, [\textrm{km}]$ along the $x$ and $z$ directions, $4\, [\textrm{km}]$ along the $y$ direction, and $0.01 \, [\textrm{m/s}]$ in each velocity direction.
This covariance is designed to be non-isotropic and to highlight the significant non-Gaussianity of the propagated distribution.
This non-Gaussianity is evident in Fig.~\ref{fig:scatter}, which shows the $x$-$y$ marginal distribution at perilune, one half period after the initial distribution epoch.
All of the Monte Carlo-based metrics in this paper are computed using 10,000 samples propagated with the Circular Restricted Three-Body dynamics.

A number of orbits in the Earth-Moon Circular Restricted Three-Body problem were considered in a similar study using the Tensor Eigenpair Measure of Nonlinearity (TEMoN) as a flag for when uncertainty propagation leads to non-Gaussianity of uncertainty \cite{gutierrez2024classifying}. Our work has a similar purpose but each method directly incorporates initial uncertainty and function nonlinearity to provide wholistic metrics rather than correlating distribution independent measures of nonlinearity with measures of non-Gaussianity to try to use one as a proxy for the other.

\subsection{Results}
\label{sec:results}
In all of the following results for each metric, we see that each measure of the quality of the linear covariance propagation degrades significantly for propagation from apolune to near perilune.
This observation agrees with previous findings \cite{jenson2023semianalytical,kulik2024applications}.
For all of the metrics besides the skewness and kurtosis, we see that the second-order, unscented transform, and Monte Carlo versions of the metrics where applicable agree very well except for potentially when the metrics are very small and sampling error for the Monte Carlo is substantial in a relative sense.
\begin{figure}[!ht]
  \centering
  \setlength\figureheight{5cm}
  \setlength\figurewidth{.7\linewidth}%
  
  \ifimporttikz
    \tikzsetnextfilename{expected_mh}
    \import{Figures/}{expected_mh.tikz}
  \else
    \includegraphics{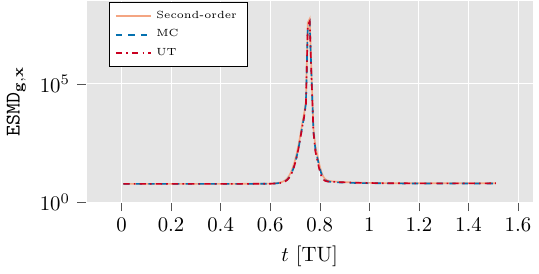}
  \fi

  \caption{The expected squared Mahalanobis distance relative to the nonlinearly propagated mean and linear covariance.}%
  \label{fig:expected_mh}
\end{figure}
\begin{figure}[!ht]
  \centering
  \setlength\figureheight{5cm}
  \setlength\figurewidth{.7\linewidth}%
  
  \ifimporttikz
    \tikzsetnextfilename{expected_error_mh}
    \import{Figures/}{expected_error_mh.tikz}
  \else
    \includegraphics{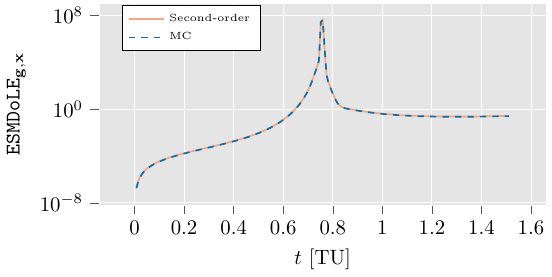}
  \fi

  \caption{The expected squared Mahalanobis distance of the linearization error with respect to the linear covariance.}%
  \label{fig:expected_error_mh}
\end{figure}
\begin{figure}[!ht]
  \centering
  \setlength\figureheight{5cm}
  \setlength\figurewidth{.7\linewidth}
  
  \ifimporttikz
    \tikzsetnextfilename{mh_mean_dif}
    \import{Figures/}{mh_mean_dif.tikz}
  \else
    \includegraphics{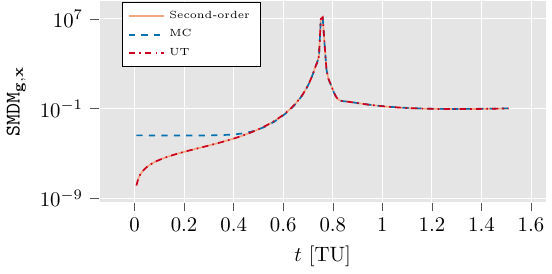}
  \fi

  \caption{The squared Mahalanobis distance of the nonlinearly propagated mean and another higher-fidelity approximation of the mean.}%
  \label{fig:mh_mean_dif}
\end{figure}
In Fig.~\ref{fig:expected_mh}, the expected squared Mahalanobis distance begins at a value of six.
Recall that six is the mean of a chi-square distribution with six degrees of freedom and the squared Mahalanobis distance of the initial distribution follows a chi-square distribution with six degrees of freedom.
Except between 0.6 and 0.9 nondimensional time units, the expected squared Mahalanobis distance stays below a value of seven.

Over the same time interval, the expected square Mahalanobis distance of the linearization error in Fig.~\ref{fig:expected_error_mh} stays below unity, though it begins at zero instead of six.
The squared Mahalanobis distance of the difference in the mean propagated versus the mean of the propagated distribution is given in Fig.~\ref{fig:mh_mean_dif}.
Here, the Monte Carlo based approach does not start at zero like the analytical approach. This is because the mean of the initial samples from the original distribution will not be exactly the mean of the distribution. By the end of the period, the value settles to around 0.1 which is lower than the value around 0.25 that is approached by the expected squared Mahalanobis distance of the linearization error. This is expected since Jenson's inequality guarantees an inequality between the two metrics.
\begin{figure}[!ht]
  \centering
  \setlength\figureheight{5cm}
  \setlength\figurewidth{.7\linewidth}
  
  \ifimporttikz
    \tikzsetnextfilename{max_cov_ratio}
    \import{Figures/}{max_cov_ratio.tikz}
  \else
    \includegraphics{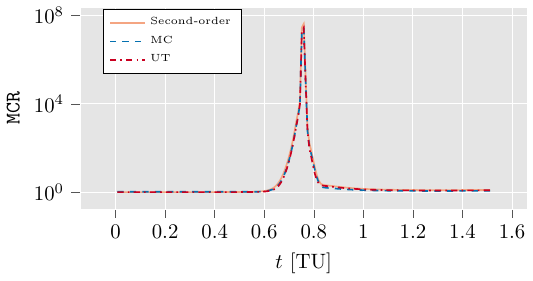}
  \fi

  \caption{The maximum ratio between the linear variance and a higher-fidelity approximation of the variance in any direction.}%
  \label{fig:max_cov_ratio}
\end{figure}

The maximum covariance ratio depicted in Fig.~\ref{fig:max_cov_ratio} begins at a value of one, settles at the end of the period to around 1.25 and 1.15 for the second-order and Monte Carlo estimates respectively.
This metric shows more disagreement between the three methods than the other expectation-based metrics.
It is also noteworthy that the shape of the graph is very similar to that of the expected squared Mahalanobis distance in Fig.~\ref{fig:expected_mh}.
This is because the trace expression in~\eqref{eq:ESMD} is equal to the sum of the generalized eigenvalues from~\eqref{eq:gen_eig} of which the maximum (or the reciprocal of the minimum) is employed as the maximum covariance ratio.
\begin{figure}[!ht]
  \centering
  \setlength\figureheight{5cm}
  \setlength\figurewidth{.7\linewidth}
  
  \ifimporttikz
    \tikzsetnextfilename{max_nonlinearity}
    \import{Figures/}{max_nonlinearity.tikz}
  \else
    \includegraphics{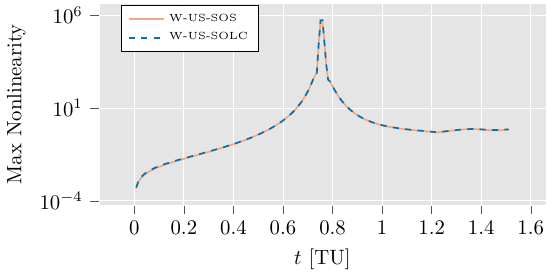}
  \fi

  \caption{The maximum uncertainty scaled nonlinearity \ac{wussos} and \ac{wussolc}.}%
  \label{fig:max_nonlinearity}
\end{figure}
\begin{figure}[!ht]
  \centering
  \setlength\figureheight{5cm}
  \setlength\figurewidth{.7\linewidth}
  
  \ifimporttikz
    \tikzsetnextfilename{max_wussadl}
    \import{Figures/}{max_wussadl.tikz}
  \else
    \includegraphics{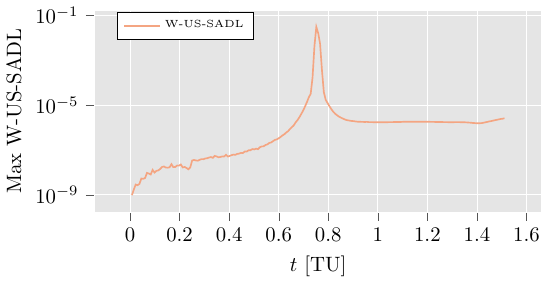}
  \fi

  \caption{The maximum uncertainty scaled nonlinearity \ac{wussadl}.}%
  \label{fig:max_wussadl}
\end{figure}

In Fig.~\ref{fig:max_nonlinearity}, both \ac{wussos} and \ac{wussolc} begin at zero and settle to a value around 0.5 by the end of a single orbit.
As noted in \cite{kulik2024applications, kulik2025NonlinearityUncertaintySplitting}, the \ac{wussolc} metric will always be greater than the \ac{wussos} metric, however in this case, the two stay consistently within a difference of a few percent between one another.
In Fig.~\ref{fig:max_wussadl}, the general trend of \ac{wussos} and \ac{wussolc} is matched but with a very different magnitude.
This discrepancy is attributed to the fact that \ac{wussadl} does not quantify a nonlinearity directly, but rather a difference in the deterministic and unscented transform-based linearizations.
At its peak, the difference between these linearizations is nearly 10\%, though past its peak, the maximum relative error between these two linearizations settles to around $10^{-6}$, indicating that the two linearizations stay very consistent with one another anywhere besides when the nonlinearity of the flow of the dynamics peaks at perilune.
While this metric is easier to compute than \ac{wussos} and \ac{wussolc} since it does not involve higher-order partial derivatives, it is less directly informative of nonlinearity.

\begin{figure}[!ht]
  \centering
  \setlength\figureheight{5cm}
  \setlength\figurewidth{.7\linewidth}
  
  \ifimporttikz
    \tikzsetnextfilename{max_skewness}
    \import{Figures/}{max_skewness.tikz}
  \else
    \includegraphics{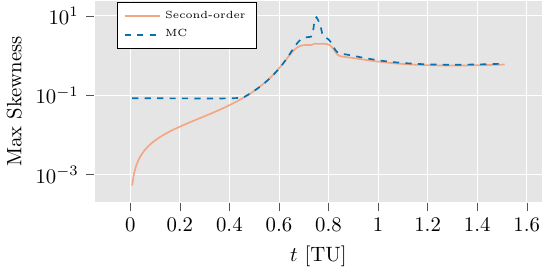}
  \fi

  \caption{The maximal normalized skewness.}%
  \label{fig:max_skewness}
\end{figure}

\begin{figure}[!ht]
  \centering
  \setlength\figureheight{5cm}
  \setlength\figurewidth{.7\linewidth}
  
  \ifimporttikz
    \tikzsetnextfilename{max_kurtosis}
    \import{Figures/}{max_kurtosis.tikz}
  \else
    \includegraphics{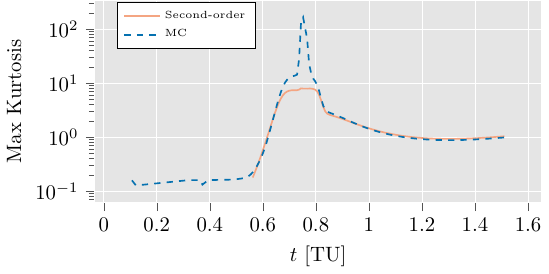}
  \fi

  \caption{The maximum excess normalized kurtosis.}%
  \label{fig:max_kurtosis}
\end{figure}

Finally, in Figs.~\ref{fig:max_skewness} and~\ref{fig:max_kurtosis}, the Monte Carlo results and the second-order analytical results are significantly different, though they both flag non-Gaussianity around the same time.
In both instances of the Monte Carlo, we see that the sample skewness and kurtosis of the initial distribution is nonzero. This effect dominates the skewness and kurtosis metrics for the Monte Carlo until around a quarter of the orbit.
Computation of the skewness and kurtosis tensors is generally more costly than the \ac{wussos} and \ac{wussolc} measures.
Thus, the \ac{wussos} and \ac{wussolc} measures offer an accurate albeit indirect indication of non-Gaussianity at a fraction of the computational cost of the direct skewness and kurtosis measures.

The additional computational penalty of the second-order kurtosis stems from additional required shifted symmetric higher-order power iterations, and, in some cases, a maximal eigenvector was not found within one thousand iterations and ten random initializations for time-of-flight below one quarter of the period.
We speculate that this is as a result of the small gap between eigenvalues of the nearly zero kurtosis tensor during that time.
Convergence of power iteration type algorithms tends to depend on the size of the difference between the largest and next largest eigenvalue.
All this goes to show that looking directly at the higher-order moments of the distribution may be the least consistent, least interpretable, and least efficient method for quantifying non-Gaussianity of those metrics proposed in this paper.

\section{Conclusion}
\label{sec:conclusion}
We have presented a number of interpretable metrics based on second-order partial derivatives and the unscented transform for assessing the error in using linear covariance propagation through nonlinear functions.
All metrics presented similar trends when employed in the context of a cislunar astrodynamics uncertainty propagation problem.
Those methods which have a Monte Carlo/sampling based equivalent (besides the maximum skewness and kurtosis characterizations), all matched the sampling based method well in the example considered.
As the scale of initial uncertainty increases, second-order models of the nonlinear function may not accurately capture the true sampling based equivalents, but should still function as an appropriate warning when linear covariance analysis is failing to accurately describe the true distribution.
These second-order validation methods are light-weight when compared with Monte Carlo-based approaches especially if the second-order partial derivatives can be calculated efficiently or precomputed and accessed efficiently, and the unscented transform based-methods are always efficient though potentially less efficient than the second-order methods when the second-order partials are already available.
Finally, we have seen that of the methods discussed, the higher-order moment-based approaches have tended to be orders of magnitude slower to compute as compared with the other methods discussed.
Additionally, the second-order approximation is not sufficient to match the Monte Carlo approach well, and third-order partial derivatives of the nonlinear function may be necessary for skewness analysis while fourth-order partial derivatives may be necessary for kurtosis analysis.
While we initially considered higher-order moment analysis as one of the most natural directions to determine non-Gaussianity and thus ineffectiveness of linear covariance analysis, without significant advances, we believe these to be less effective methods than those that more directly consider uncertainty-weighted measures of nonlinearity.

\backmatter

\section*{Funding Declaration}
Part of this work was sponsored by the United States Air Force Research Laboratory and the United States AFRL Regional Hub and was accomplished under Cooperative Agreement Number FA8750-22-2-0501. The views and conclusions contained in this document are those of the authors and should not be interpreted as representing the official policies, either expressed or implied, of the United States Air Force or the U.S. Government. The U.S.  Government is authorized to reproduce and distribute reprints for Government purposes notwithstanding any copyright notation herein.

\bibliographystyle{unsrt}
\bibliography{sn-bibliography}%

\end{document}